\documentclass[12pt]{article}
\usepackage{amsmath}
\usepackage{graphicx}
\usepackage{enumerate}
\usepackage{natbib}
\usepackage{url} 

\usepackage{amsthm}
\usepackage{algorithm}
\usepackage{algpseudocode}
\usepackage[english]{babel}
\usepackage{bm}
\usepackage{bbm}
\usepackage[]{color}
\usepackage[mathscr]{eucal}
\usepackage{fancyhdr}
\usepackage{float}
\usepackage[T1]{fontenc}
\usepackage{hyperref}
\usepackage[utf8]{inputenc}
\usepackage{mathtools}
\usepackage{neuralnetwork}
\usepackage{setspace}
\usepackage{subfigure}
\usepackage{tikz}
\usepackage{titling}
\usepackage{adjustbox}
\usepackage{multirow}
\usepackage{amsfonts}
\usepackage{soul}
\usepackage{listofitems} 
\usetikzlibrary{arrows.meta} 
\usepackage[outline]{contour} 
\contourlength{1.4pt}

\newcommand{\blind}{1}

\renewcommand{\hat}[1]{\widehat{#1}}

\newtheorem{theorem}{Theorem}
\DeclarePairedDelimiter{\norm}{\lVert}{\rVert}

\tikzset{>=latex} 
\usepackage{xcolor}
\colorlet{myred}{red!80!black}
\colorlet{myblue}{blue!80!black}
\colorlet{mygreen}{green!60!black}
\colorlet{myorange}{orange!70!red!60!black}
\colorlet{mydarkred}{red!30!black}
\colorlet{mydarkblue}{blue!40!black}
\colorlet{mydarkgreen}{green!30!black}
\tikzset{
  >=latex, 
  node/.style={thick,circle,draw=black,minimum size=22,inner sep=0.5,outer sep=0.6},
  node in/.style={node,draw=black},
  node hidden/.style={node,draw=black},
  node hiddenellip/.style={draw=none, fill=none},
  node convol/.style={node,draw=black},
  node out/.style={node,draw=black},
  connect/.style={thick,mydarkblue}, 
  connect arrow/.style={-{Latex[length=4,width=3.5]},thick,shorten <=0.5,shorten >=1},
  node 1/.style={node in}, 
  node 2/.style={node hidden},
  node 3/.style={node out}
}

\begin{document}

\def\spacingset#1{\renewcommand{\baselinestretch}%
{#1}\small\normalsize} \spacingset{1}


\if1\blind
{
  \title{\bf Likelihood-free Posterior Density Learning for Uncertainty Quantification in Inference Problems}
  \author{Rui Zhang 
    Department of Statistics, The Ohio State University\\
    and \\
    Oksana A Chkrebtii \\
    Department of Statistics, The Ohio State University}
  \maketitle
} \fi

\if0\blind
{
  \bigskip
  \bigskip
  \bigskip
  \begin{center}
    {\LARGE\bf Likelihood-free Posterior Density Learning for Uncertainty Quantification in Inference Problems}
\end{center}
  \medskip
} \fi

\begin{abstract}
Generative models and those with computationally intractable likelihoods are widely used to describe complex systems in the natural sciences, social sciences, and engineering. Fitting these models to data requires likelihood-free inference methods that explore the parameter space without explicit likelihood evaluations, relying instead on sequential simulation, which comes at the cost of computational efficiency and extensive tuning. We develop an alternative framework called kernel-adaptive synthetic posterior estimation (KASPE) that uses deep learning to directly reconstruct the mapping between the observed data and a finite-dimensional parametric representation of the posterior distribution, trained on a large number of simulated datasets. We provide theoretical justification for KASPE and a formal connection to the likelihood-based approach of expectation propagation. Simulation experiments demonstrate KASPE's flexibility and performance relative to existing likelihood-free methods including approximate Bayesian computation in challenging inferential settings involving posteriors with heavy tails, multiple local modes, and over the parameters of a nonlinear dynamical system.

\end{abstract}

\noindent%
\vfill

\newpage
\spacingset{1.9} 

\section{Introduction}\label{Introduction}
A fast, accurate, and general approach for Bayesian inference without access to a likelihood is still very much an open area of research. While exact methods exist for specific settings and model forms \citep{Andrieu2009,Atchadé2013}, more general approaches typically require some degree of approximation \citep{Fearnhead2012} as well as tuning and expensive computation. Likelihood-free problems occur in many settings, such as when generative models emulate complex systems, from anthropology \citep[e.g.,][]{Cegielski2016}, to ecology \citep[e.g.,][]{Chkrebtii2014}, to systems biology \citep[e.g.,][]{Boys2008}. Here a clear, and often simple set of rules is built into a stochastic system which can then be simulated forward in time, but a closed-form likelihood is unavailable. Likelihoods can be intractable when they contain combinatorially large numbers of components arising in models such as interacting atomic spins on lattices \citep[e.g.,][]{Ghosal2020, Atchadé2013} and networks \citep[e.g.,][]{Stivala2020}, or an intractable normalizing constant, such as probability models defined on a manifold \citep[e.g.,][]{Fallaize2016}, and Gaussian random fields \citep[e.g.,][]{Varin2011}; or depend on latent variables, such as in state space models \citep[e.g.,][]{Durbin2012}, hidden Markov models \citep[e.g.,][]{Yildirim2015}, and mixed and random effects models \citep[e.g.,][]{Varin2011}, where the likelihood is a high-dimensional integral or summation over all latent variable values.

A popular approach for Bayesian likelihood-free inference is approximate Bayesian computation (ABC) \citep{Tavare1997, Pritchard1999, Beaumont2002}, consisting of sampling techniques that target an approximate posterior distribution \citep{Fearnhead2012}, termed the ABC posterior, obtained by replacing the likelihood with a kernel density approximation based on the discrepancy between summarized synthetic and observed data. Despite its generality, ABC is both computationally expensive and requires careful tuning and convergence monitoring.

Neural networks (NN) have recently shown promise for likelihood-free parameter estimation, despite practical drawbacks. For example, \cite{lenzi2021} and \cite{SainsburyDale2024} propose to reconstruct the mapping from the data space to the parameter space by training a NN model on a large number of simulated data-parameter pairs. \cite{Zhang2024} formalizes the method, provides a connection to Bayes estimation, and introduces necessary dimension reduction. Bayesian NN-methods are based on learning the posterior density from synthetic training data. Mixture density networks (MDNs; \citealp{bishop1994mdn}) estimate the parameters defining a Gaussian mixture that approximates the posterior distribution, but suffer from substantial bias in realistic scenarios where the prior is diffuse and training data are sparse. Variant of the MDN framework introduced by \cite{Papamakarios2016,lueckmann2017flexible} nonetheless suffer from numerical instability, lack of generality, requiring multiple rounds of neural network training at significant computational cost. 
Unlike methods that employ a calibration kernel for importance weighting in multi-round sequential neural posterior estimation (e.g., \citealp{lueckmann2017flexible}; \citealp{Xiong2026}), our approach leverages kernel-weighted synthetic training data within a single-stage, non-sequential framework to directly learn the posterior mapping, thereby stabilizing learning and reducing estimation variability.
Beyond MDN-based approaches, recent advances incorporate normalizing flows to increase the flexibility of posterior approximations, but require likelihood evaluation. Normalizing flows \citep{rezende2015variational, papamakarios2019normalizing} are a technique for estimating densities by transforming a simple base distribution into a target distribution through a sequence of invertible transformations with tractable Jacobians, often parameterized by NNs. Variational inference with normalizing flows \citep{kingma2016improving} enhances the expressiveness of variational approximations, but requires likelihood evaluation during training to optimize the flow parameters and the variational objective, rendering it inapplicable in simulation-based settings. Likelihood-free adaptations, such as conditional invertible neural networks (cINNs; \citealp{winkler2019conditional, ardizzone2019guided}), learn an invertible mapping between parameters and latent variables conditioned on observed data, enabling efficient posterior sampling. However, the requirement of global invertibility can restrict architectural flexibility and complicate training, making these models sensitive to initialization and challenging to tune in practice. An alternative is the automatic posterior transformation (APT) approach proposed by \cite{greenberg2019automatic}, which recasts inference as a density ratio estimation problem. It leverages flow-based density estimators and dynamically updated proposals to flexibly approximate the posterior, but still requires multiple rounds of training.

We develop an alternative likelihood-free inference method called kernel-adaptive synthetic posterior estimation (KASPE), which learns a parametric approximation to the exact posterior via deep learning combined with a kernel-based adaptive sampling mechanism to generate synthetic training data. We demonstrate that this approach has connections with the likelihood-based method of expectation propagation, and contains the likelihood-free MDN estimation method as a special, but less efficient, case.

The remainder of this paper is structured as follows: Section \ref{ch3.background} provides a background on ABC and EP methods, discussing their advantages and drawbacks. Section \ref{ch3.method} details the methodology of KASPE and presents its connection to the EP and MDN. Section \ref{ch3.examples} presents numerical experimental results and discusses the performance improvements achieved through our proposed approach. Finally, Section \ref{ch3.discussion} concludes the paper with a summary of findings.

\section{Background}\label{ch3.background}
Consider data $y_0\in \mathbb{R}^m$ generated from a probability model known up to some parameters $\theta\in \Theta\subseteq \mathbb{R}^d$, with (possibly unknown) likelihood function denoted by $p(y \mid \theta)$. 
In particular, we assume that synthetic data $y$ can be readily simulated from $p(y \mid \theta)$ for arbitrary values of $\theta\in \Theta$, but that the likelihood is computationally inaccessible. We wish to make posterior inference on $\theta$ given the observed data $y_0$ and prior density $\pi(\theta)$, but without access to $p(y\mid \theta)$. 
In this section we review the likelihood-free Bayesian parameter estimation approach of ABC, and the likelihood-based approach of EP, and discuss their features and limitations.

\subsection{Approximate Bayesian Computation}\label{ABC-section}
ABC is a class of sampling techniques that targets the approximate ABC posterior, 
\begin{equation*}
       \pi_{\text{ABC}}(\theta \mid s_0)\propto \pi(\theta)\int p(y \mid \theta)K\left(\frac{S(y)-s_0}{h}\right) dy
\end{equation*}
where  $s_0=S(y_0)$ is a summary of the observed data, $K(\cdot)$ is a kernel function that integrates to 1, with $\max\{K(x)\}=1$, and $h>0$ is a user-selected bandwidth parameter. 
Inference is based on MCMC estimates of the functionals of the ABC posterior. As an illustration, the ABC Markov chain Monte Carlo (ABC-MCMC) sampler is presented in Algorithm \ref{ABC-MCMC}. 
\begin{algorithm}
\caption{Algorithm for ABC-MCMC}\label{ABC-MCMC}
\begin{algorithmic}[1]
    \Require observed data $y_0, \text{ likelihood function }p(\cdot \mid \cdot), \text{ prior density } \pi(\cdot),$  \Statex $\text{ summary function } S(\cdot),
     \text{ kernel function } K(\cdot), \text{ proposal density }g(\cdot \mid \cdot),$
     \Statex $\text{ bandwidth } h>0, \text{ integer } n>0$ 
\Ensure ABC posterior samples  $\{\theta_i\}_{i=1}^n$
\State initialize $\theta_c$ and sample $y_c\sim p(\cdot \mid \theta_c)$ 
\State define $s_0=S(y_0)$ and $s_c=S(y_c)$
\For{$i=1$ to $n$}
\State sample $\theta \sim g(\cdot \mid \theta_c)$
\State sample $y \mid \theta \sim p(\cdot \mid \theta)$
\State define $s=S(y)$
\State with probability 
$$\text{min}\left(1, \frac{K(\frac{s-s_0}{h})}{K(\frac{s_c-s_0}{h})}\frac{\pi(\theta)}{\pi(\theta_c)}\frac{g(\theta_c \mid \theta)}{g(\theta \mid \theta_c)}\right)$$
accept $\theta$ and set $\theta_c=\theta, s_c=s$; otherwise, keep $\theta_c \text{ and }s_c$ unchanged
\State $\theta_n=\theta_c$
\EndFor
\State obtain samples $\{\theta_i\}_{i=1}^n$ from ABC posterior
\end{algorithmic}
\end{algorithm}

Although the approximation of the posterior improves as the dimension of the summary statistic grows, the Monte Carlo error increases as the probability of accepting sufficiently many simulations decreases. The choice of summary statistics is problem-dependent and must take into account this trade-off. A similar trade-off exists with the choice of the bandwidth parameter. Moreover, ABC algorithms cannot be fully parallelized which, combined with typically low acceptance probabilities, result in potentially very long run-times.

\subsection{Expectation Propagation}\label{VI}
The \emph{expectation propagation} (EP) approach \citep{minka2001expectation} casts inference as an optimization problem over a class of densities in order to find the one that is closest to the posterior. Denote by $p(y)$ the marginal density of $y$ with support $\mathcal{Y}$. The user chooses a family of densities $\mathcal{Q}=\{q(\cdot \mid \eta): \eta\in \mathcal{E}\}$ over $\theta$ parameterized by auxiliary variables $\eta\in \mathcal{E}\subseteq \mathbb{R}^v$. Given observations \( y_0 \), EP seeks the parameterization \( \eta(y_0) \in \mathcal{E} \) that produces the distribution \( q \) closest to the posterior density of interest \( \pi(\cdot \mid y_0) \), where the discrepancy is measured by the Kullback–Leibler (KL) divergence. This is formalized as the  optimization problem:
\begin{equation*}
\eta(y_0) = \underset{\eta \in \mathcal{E}}{\arg \min} \ \text{KL}\left(\pi(\cdot \mid y_0) \mid q(\cdot \mid \eta)\right).
\end{equation*}
This minimization is typically performed using numerical optimization algorithms, and the estimated parametrization is not necessarily unique. The resulting estimate of the posterior density is $\hat{\pi}^{EP}(\cdot \mid y_0)=q(\cdot \mid \eta(y_0))$ before the application of the mean-field assumption, which will be discussed shortly. $\text{KL}\left(\pi(\cdot \mid y_0) \mid  q(\cdot \mid \eta)\right)$ is the KL divergence from $\pi(\cdot \mid y_0)$ to $q(\cdot \mid \eta)$, defined as
\begin{align*}
    \text{KL}\left(\pi(\cdot \mid y_0) \mid  q(\cdot \mid \eta)\right):&=\mathbb{E}_{\theta\sim \pi(\theta \mid y_0)}\left(\log \frac{\pi(\theta \mid y_0)}{q(\theta \mid \eta)}\right).
\end{align*}
It can be shown that if the posterior density is contained in the family of probability density functions, then the EP estimate without the mean-field assumption is equal to the true posterior density except on a set of Lebesgue measure zero, since $\text{KL}\left(\pi(\cdot \mid y_0) \mid  q(\cdot \mid \eta)\right)$ attains its global minimum of zero if and only if the two distributions are identical almost everywhere.
In general, however, this minimization will be intractable because the KL divergence
involves an integral with respect to the true posterior distribution. Therefore, in practice, EP approximates the posterior distribution $\pi(\cdot \mid y_0)$ by a product of independent factors, known as the \emph{mean-field assumption}. This assumption simplifies the optimization problem by breaking down the complex posterior into more manageable components. If, further, the family of densities $q(\cdot \mid \eta)$ is a product of independent factors from a chosen distribution family, then the EP posterior can be estimated efficiently by iteratively updating the factors to minimize the KL divergence between the true posterior distribution and the approximate posterior distribution. This iterative process involves passing messages between the factors, with each factor being updated based on the messages received from the other factors. The updates are done in such a way that the KL divergence decreases at each iteration, converging to a local minimum.
EP is often easier to implement compared to sampling-based Bayesian methods like Markov Chain Monte Carlo (MCMC), as it does not require careful tuning, making it more accessible to practitioners. 

Critically to our likelihood-free inference setting, EP relies on access to a tractable likelihood and lacks generality, requiring non-trivial analytical work to derive factor updates. Another important limitation of EP is that there is no guarantee that the iterations will converge, which can be due to the sensitivity to initial values, model complexity, and data characteristics leading to multiple local minima in the optimization landscape \citep{minka2001expectation}. Further, EP may face scalability issues in very high-dimensional or complex models, as the number of local updates and computations required can become prohibitive. Additionally, the assumption that the posterior distribution can be approximated by a factorized form does not always hold, leading to poor approximations. In fact, EP produces poor approximations if the posterior distribution is multi-modal: the algorithm tends to average over the different modes, leading to a single-mode approximation that significantly distorts the representation of the true multimodal posterior \citep{bishop2006pattern}.

\section{Methods}\label{ch3.method}

\emph{Kernel-adaptive synthetic posterior estimation} (KASPE) is a likelihood-free posterior estimation approach which uses synthetic data drawn from the data-generating mechanism to learn a deep NN which maps the input $y$ to the parameters $\eta$ defining a parametric model that best approximates the posterior density in terms of a KL divergence-based loss. A kernel-based sampling mechanism selectively accepts synthetic data based on a kernel-defined probability measure of the discrepancy between synthetic and observed data, prioritizing samples whose outputs align with the observations.  
This section introduces KASPE and explores its connection with the existing approaches of \textit{expectation propagation} (EP) and \textit{mixture density network} (MDN) estimation. A discussion of optimal algorithm settings completes the section.

\subsection{Kernel-adaptive Synthetic Posterior Estimation}\label{KASPE intro}
KASPE consists of two steps: kernel-weighted synthetic training simulation, followed by a neural network (NN) reconstruction of the mapping between the data $y_0$ and a finite-dimensional representation of the posterior density $q$. As in EP, the candidate family of densities is taken to be $\mathcal{Q}=\{q(\cdot \mid \eta): \eta\in \mathcal{E}\}$. The specific choice of densities is discussed in Section \ref{sec:settings}.

Let $x$ be a $m$-dimensional input vector and define the kernel function $K(x)$, which is non-increasing with respect to the norm $\| x\|$, integrates to 1, and has $\max\{K(x)\}=K(0)=1$. $N$ synthetic training data-parameter pairs $(\theta_{i},y_{i})_{i=1}^{n}$ are obtained by first sampling the parameter from its prior distribution and then simulating the data from the model given the parameter, i.e., $(\theta_{i},y_{i})\stackrel{\text{ind}}{\sim}\pi(\theta)p(y \mid \theta)$ where $y_i$ has the same dimension as the observed data. The data-parameter pair $(\theta_{i},y_{i})$ is retained with probability $K(\frac{y_i-y_0}{h})$, and discarded otherwise, i.e., we introduce a weight $w_i$ associated with the data-parameter pair $(\theta_{i},y_{i})$ such that $w_i=1$ with probability $K(\frac{y_i-y_0}{h})$, and $w_i=0$ otherwise. The collection of the $n$ synthetic training data-parameter pairs and their  weights, denoted by $\{(\theta_{i}, y_{i}, w_i)_{i=1}^{n}\}$, will be referred to as \emph{synthetic training data}.
The \emph{effective sample size} $n_\text{eff}$ is defined as the sum of the realized weights: $n_\text{eff}=\sum_{i=1}^{n}w_i$, corresponding to the total number of accepted proposals. 
Intuitively, the kernel function measures the closeness of the simulated and observed data, so that parameter proposals that produce $y_n$ close to the observed data are more likely to be accepted. 
The adaptation provided by the kernel-based weight serves to guide samples toward regions of higher posterior probability, offering an effective means of reducing Monte Carlo approximation error, since in most cases the parameter space is considerably larger than the region of highest posterior density, and the prior distribution is substantially more diffuse than the posterior distribution.

For a given NN architecture, we define the vector-valued \emph{NN function} $\mathbf{N}(\cdot,\omega):\mathbb{R}^m \rightarrow \mathcal{E}$, where vector $\omega \in \Omega$ denotes the NN weight parameters. The input is data $y\in \mathcal{Y}\subseteq \mathbb{R}^m$, where $\mathcal{Y}$ denotes the support of $y$, and the output consists of auxiliary parameter vector $\eta\in \mathcal{E}$ indexing $q$, as illustrated in Supplement Figure S.1. A weighted negative log posterior density $-w\log q(\theta \mid \mathbf{N}(y,\omega))$ defines the training loss function $Q_n(\omega) = -\frac{1}{n} \sum_{i=1}^{n} w_i\log q(\theta_{i} \mid \mathbf{N}(y_{i},\omega))$,  
which penalizes NN weights that lead to small average training data densities given their simulation parameters. 
The KASPE estimate of the posterior density is the parametric function $q$ indexed by parameters $\eta$ obtained by mapping observed data $y_0$ to $\mathcal{E}$ through a NN trained on the synthetic data, i.e.
\begin{equation}
\hat{\pi}_n(\cdot \mid y_0)=q(\cdot \mid \mathbf{N}(y_0, \hat{\omega}_n)), \; \text{ where } \hat{\omega}_n\in \underset{\omega\in\Omega}{\arg \min }\ -\frac{1}{n} \sum_{i=1}^{n}w_{i}\log q(\theta_{i} \mid \mathbf{N}(y_{i},\omega)).  \label{KASPE est}
\end{equation} 

As with other popular likelihood-free methods, such as ABC, the quality of the KASPE estimator degrades quickly with the data dimension.  To overcome this effect, we summarize the data via lower-dimensional statistics that are, ideally, as close to sufficient as possible, though this cannot be guaranteed in general for the likelihood-free setting.    
Let $S(\cdot):\mathbb{R}^{m}\rightarrow \mathbb{R}^{K}$ be a function which summarizes the $m$-dimensional data with a $K$-dimensional summary statistic. 
We replace $y$ by $s$ in both kernel-based proposal mechanism and NN training, so that distance will be measured in terms of summaries and the NN will take summary statistics as inputs. The dimension-reduced method, KASPE-DR, is provided in Algorithm \ref{KASPE-DR algo} and contains KASPE as a special case when the data summary is the identity. 

\begin{algorithm}
\caption{Algorithm for KASPE-DR}\label{KASPE-DR algo}
\begin{algorithmic}[1]
    \Require observed data $y_0, \text{ likelihood function }p(\cdot \mid \cdot), \text{ prior density } \pi(\cdot), $\Statex $\text{ summary function } S(\cdot), \text{ family of distributions } q(\cdot \mid \cdot), \text{ NN function }\mathbf{N}(\cdot,\cdot), $
    \Statex $ \text{ bandwidth } h>0, \text{ integer } n>0$ 
\Ensure $\hat{\pi}_n(\cdot \mid y_0)$
\For{\textcolor{red}{$i=1$ to $n$}}
\State sample $\theta_{i}\sim \pi(\cdot)$ \label{line:sample_theta}
\State sample $y_{i} \mid \theta_{i} \sim p(\cdot \mid \theta_{i})$
\State let $s_0=S(y_0)$ and $s_i=S(y_i)$ for $i=1,\ldots, n$
\State with probability $K(\frac{s_i-s_0}{h})$, set $w_i=1$; otherwise, we set $w_i=0$
\EndFor
\State use numerical optimization to solve $$\hat{\omega}_n\in\underset{\omega\in\Omega}{\arg \min }\ -\frac{1}{n} \sum_{i=1}^{n}w_{i}\log q(\theta_{i} \mid \mathbf{N}(s_{i},\omega))$$
\State set posterior density estimate $\hat{\pi}^{DR}_n(\cdot \mid s_0)=q(\cdot \mid \mathbf{N}(s_0,\hat{\omega}_n))$
\end{algorithmic}
\end{algorithm}

In order to avoid NN overfitting in both MDN or KASPE methods, we generate \textit{synthetic validation data} in the same manner as the remaining training data. The optimization algorithm's stopping time (maximum number of epochs), is then determined by minimizing the validation loss rather than the training loss.

\subsection{Algorithm Settings}\label{sec:settings}
The choice of the family of densities is application-specific. However, a default suggestion is to use a mixture of multivariate normal densities $\phi_l$ with free mean and covariance $\mu_l$ and $\Sigma_l$, respectively. 
To ensure positive definiteness of the estimated covariance matrices, we parameterize $\phi_l$ by the Cholesky factor $U_l$ of the covariance inverse, i.e.,  $\Sigma_l^{-1}=U_l^\top U_l$, where $U_l$ is an upper triangular matrix with strictly positive diagonal elements, and  $|\Sigma_l^{-1}|=\left(\prod_{k=1}^{d} \text{diag}(U_l)_k\right)^2$, where $\text{diag}(U_l)_k$ denotes the $k$-th diagonal element of $U_l$. 
Thus, the corresponding candidate family of densities is,
$$
q(\theta \mid \eta) =\sum_{l=1}^{L}\alpha_l\phi_l(\theta \mid \mu_l,U_l),$$
where, 
$$\phi_l(\theta \mid \mu_l,U_l)=(2 \pi)^{-d/2} \prod_{k=1}^{d} \text{diag}(U_l)_k \exp \left\{-\frac{1}{2}\norm{U_l(\theta-\mu_l)}^2\right\}.  \label{KASPE mixture of density}$$
This parametrization has $\frac{1}{2}L(d+1)(d+2)$ auxiliary parameters.

The activation functions for the nodes in the output layer of the neural network must be chosen in such a way as to satisfy any required parameter restrictions. Let $\mu_{lj}$ be the $j$th element of the vector $\mu_l$, and let $(U_l)_{jk}$ be the $(j,k)$th entry of the matrix $U_l$.
Denote by $z_{l}^{\alpha}$, $z_{lj}^{\mu}$, and $z_{ljk}^{U}$ the neural network outputs corresponding to $\alpha_l$, $\mu_{lj}$, and $(U_l)_{jk}$, respectively,  before an activation function is applied.  
The sum-to-one constraint on the mixing coefficients $\alpha_l$ can be achieved by using a softmax activation function $\alpha_l=\frac{\exp(z_{l}^{\alpha})}{\sum_{l=1}^{L}\exp(z_{l}^{\alpha})}$. 
For the upper triangular matrix $U_l$ with strictly positive elements in the
diagonal, an exponential activation function is applied to diagonal elements to enforce positivity, namely, 
$$ (U_l)_{jk}= \exp (z_{ljk}^{U}) \mathbf{1} \{j = k\} + z_{ljk}^{U} \mathbf{1}\{j < k \}.$$
Since the multivariate mean parameters $\mu_{lj}$ are unconstrained, a linear activation function $\mu_{lj}=z_{lj}^{\mu}$ may be used for the corresponding nodes. 

Another important setting for implementing KASPE is the tuning parameter $h$. 
Although we would like the bandwidth parameter to be as small as possible, we must balance this choice with the need for a suitable acceptance rate for synthetic training samples. So to choose an appropriate bandwidth parameter, we suggest performing a pilot run by first generating $N$ synthetic data-parameter pairs, and deriving an estimate of the expected acceptance rate as a function of $h$, namely, $a(h)= N^{-1} \sum_{n=1}^{N}K(\frac{y_n-y_0}{h})$. Under a fixed computational budget, we can specify a lowest acceptable expected acceptance rate and use this to find the corresponding value of $h$.

\subsection{Properties and Connection with Existing Approaches}\label{property}
In this section, we establish a connection between the KASPE estimator and the EP method, and show that the KASPE estimator is consistent under some conditions. We further show that MDN is a special case of KASPE with constant weight.  For a given NN architecture, denote the space of vector-valued NN functions as $\mathcal{M}=\{\mathbf{N}(\cdot,\omega) \mid \omega\in \Omega\}$, where $\mathbf{N}(\cdot,\omega):\mathbb{R}^m \rightarrow \mathcal{E}$. Note that \( Q_n(\omega) \) is a random function of \( \omega \), where randomness is induced by the synthetic training data  \( (\theta_i, y_i, w_i) \).

Define the expected training loss function as $$Q_0(\omega)=\mathbb{E}_{(\theta, y, w)}\bigl[Q_n(\omega)\bigr]= -\mathbb{E}_{(\theta, y, w)}\bigl[w\log q(\theta \mid \mathbf{N}(y,\omega))\bigr]=-\mathbb{E}_{(\theta, y)}\left[K\left(\frac{y-y_0}{h}\right)\log q(\theta \mid \mathbf{N}(y,\omega))\right],$$ and the KASPE parameterization estimator as  
\[
\hat{\eta}_n(\cdot) := \mathbf{N}(\cdot, \hat{\omega}_n), \text{ where }  \hat{\omega}_n \in \underset{\omega \in \Omega}{\operatorname{argmin}} \; Q_n(\omega).
\]
The following theorems, proved in the Supplement, formalize the connection between KASPE and EP estimation.

\begin{theorem}\label{consistency}
Assume that:
\begin{enumerate}
    \item The parameter space $\Omega$ of the NN weights is compact;
    \item The NN function $\mathbf{N}(y, \omega)$ is continuous in $\omega$ for any fixed $y\in \mathcal{Y}$;
    \item The expected training loss function $Q_0(\omega)< \infty$ for any $\omega\in \Omega$ and has a set of minimizers $\Omega_0 = \underset{\omega\in\Omega}{\operatorname{argmin}} \, Q_0(\omega)$ that satisfies, for any \( \omega_a, \omega_b \in \Omega_0 \), 
    \(
        \mathbf{N}(\cdot, \omega_a) = \mathbf{N}(\cdot, \omega_b).
    \)
    That is, the induced NN function at the minimizers is unique, denoted as $\mathbf{N}_0(\cdot)$;
    \item The training loss function converges to the expected training loss function uniformly in probability: $\sup_{\omega \in \Omega} |Q_n(\omega) - Q_0(\omega)| \xrightarrow{p} 0$ as $n\to\infty$.
\end{enumerate}
Then, the KASPE parameterization estimator $\hat{\eta}_n(\cdot)$ converges pointwise in probability to the function $\mathbf{N}_0(\cdot)$ as the number of synthetic weighted training data-parameter pairs $n\to\infty$. That is, for each fixed $y\in \mathcal{Y}$:
\[
    \hat{\eta}_n(y)\xrightarrow{p} \mathbf{N}_0(y), \text{ as }n \to \infty.
\]
\end{theorem}

\begin{theorem}\label{convergence}
Suppose there exists an EP parameterization estimator $\eta(\cdot)$ within $\mathcal{M} = \{\mathbf{N}(\cdot,\omega) : \omega \in \Omega\}$, and Assumptions 1--4 in Theorem \ref{consistency} hold. Then the KASPE parameterization estimator $\hat{\eta}_n(\cdot)$ converges pointwise in probability to the EP parameterization estimator $\eta(\cdot)$ as the number of synthetic weighted training data-parameter pairs $n\to\infty$. That is, for each fixed $y\in \mathcal{Y}$:
\[
    \hat{\eta}_n(y)\xrightarrow{p} \eta(y), \text{ as }n \to \infty.
\]
\end{theorem}

Theorem \ref{convergence} shows that if the neural network model space $\mathcal{M}$ is sufficiently rich, then under mild conditions, for any $y \in \mathcal{Y}$, the KASPE estimator of the posterior density converges in probability to the EP estimate obtained  without the mean-field assumption as the number of synthetic weighted training data-parameter pairs grows. Furthermore, as discussed in Section \ref{VI}, if the EP family of densities is sufficiently rich to contain the true posterior density, the KASPE estimate will converge in probability to a density equal to the true posterior density except on a set of Lebesgue measure zero. While this result establishes convergence for any fixed $y$, the use of the kernel function $K\bigl(\tfrac{y - y_0}{h}\bigr)$ prioritizes training samples with $y_i$  near the observed data $y_0$. As a result, the estimator effectively utilizes more informative samples at the target point $y = y_0$, leading to faster convergence rates and lower variance at this point, which is typically the primary interest in posterior inference.

\textit{Mixture Density Network} (MDN) estimation \citep{bishop1994mdn} is another neural network-based approach for posterior density estimation that only requires the ability to simulate from the model, rather than full likelihood evaluation. MDN estimation can be viewed as a special, sub-optimal case of the KASPE approach. The mixture density network, as implied by its name, selects a mixture of densities as its family of density functions
\begin{equation}
q(\theta \mid \eta)=\sum_{l=1}^{L}\alpha_l\phi_l(\theta \mid \eta_l),  \label{mixture of density}
\end{equation}
where $L$ is the number of components in the mixture, $\phi_l(\theta \mid \eta_l)$ are component density functions parameterized by $\eta_l$, and $\alpha_l$ are non-negative mixing coefficients that sum to $1$. For example, \cite{bishop1994mdn} chose Gaussian component density functions:
\begin{equation}
\phi_l(\theta \mid \eta_l)=\phi_l(\theta \mid \mu_l,\sigma_l) = \frac{1}{(2 \pi)^{d/2} \sigma_l^d} \exp \left\{-\frac{\left\| \theta-\mu_l\right\| ^2}{2 \sigma_l^2}\right\}, \;\; \sigma_l > 0,  \label{MDN component function}
\end{equation}
which assumes that within each component of the mixture distribution, the covariance matrix is diagonal with identical diagonal elements. Under this model choice, the total number of parameters defining the MDN is $L(d+2)$. 
In general, the MDN method accepts all randomly selected synthetic training data, making it an extreme case of KASPE with $K(x)\equiv1$.

\section{Simulation Experiments}\label{ch3.examples}
Three simulation experiments illustrate the relative performance of KASPE, approximate Bayesian computation (ABC), and mixture density network (MDN) posterior estimation. We consider posterior densities with (i) skewness and heavy tails, (ii) multiple local modes, and (iii) over parameters defining a nonlinear ordinary differential equation (ODE) model from discretely observed data. The first two models have closed-form posteriors, while the third features a tractable likelihood that enables the use of MCMC as a benchmark. For each scenario, we consider observation sample sizes $m = 4, 100, 1,000$, except for the third example, where the smallest sample size $m=6$ is chosen to obtain a multi-modal posterior with well-separated local modes. 
For the largest sample size of $m = 1,000$, we consider the general form of the three approaches, as well as their recommended  dimension-reduced versions (renamed by appending ``-DR''). Multiple runs of each algorithm for a given dataset allow us to visualize sampling variation for each method. 
For each setting in the first two examples, we provide a summary figure showing the estimated (dotted lines) and true (solid line) marginal posteriors for each approach, and heatmaps of the estimated and true joint posteriors. 
The candidate densities for KASPE and MDN are a mixture of 20 Gaussian densities with free mixture weights, mean vectors, and covariance matrices. The NN architecture is a feed-forward network with 2 hidden layers and training sample size of $n = 125,000$ where $25\%$ is held out for validation. For KASPE and ABC, a squared exponential kernel is used to measure distance between observed and synthetic data, with the bandwidth parameter chosen manually to be as small as possible while targeting the desired acceptance rate. ABC is implemented using adaptive tuning of the proposal covariance within a parallel-tempering ABC-MCMC algorithm to enable the sampler to efficiently explore posteriors with possible local modes \citep{Swendsen1986, Geyer1991}. Convergence is assessed by monitoring traceplots and correlation plots. The ABC posterior density is approximated from the ABC sample via kernel density estimation. Additional simulation settings and figures are provided in Supplement Sections S.3 and S.4.

\begin{figure}[ht]
	\begin{center}
		\subfigure{%
			
			\includegraphics[width=0.7\textwidth]{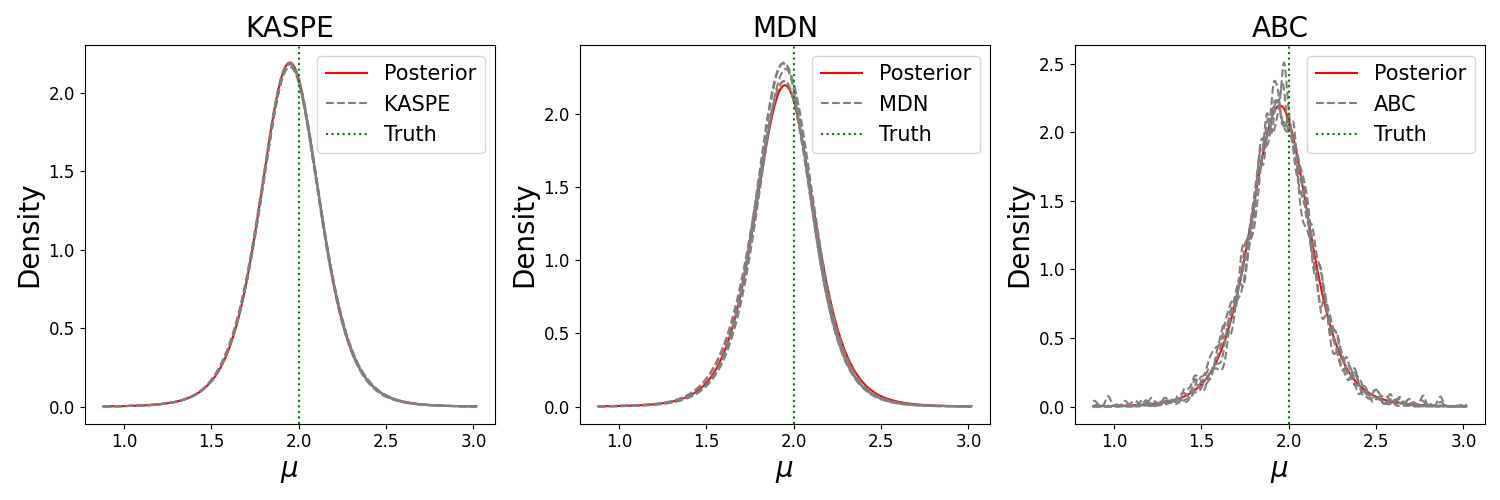}
		}\\
        \vspace{-0.2in}
		\subfigure{%
			
			\includegraphics[width=0.7\textwidth]{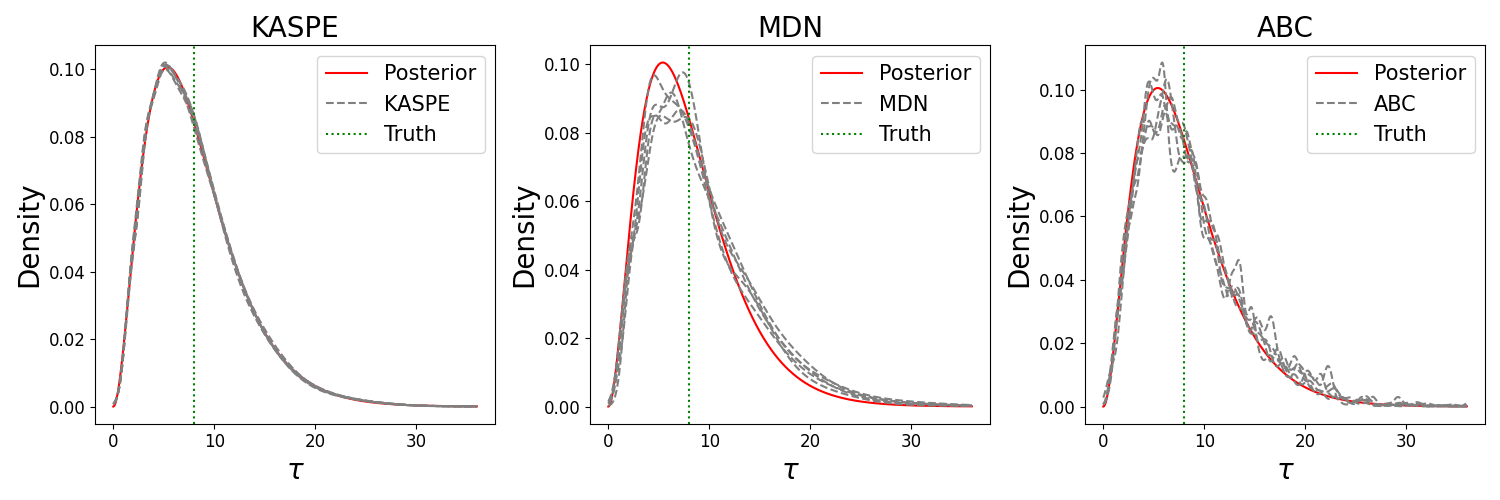}
		}\\
        \vspace{-0.2in}
		\subfigure{%
			
			\includegraphics[width=0.8\textwidth]{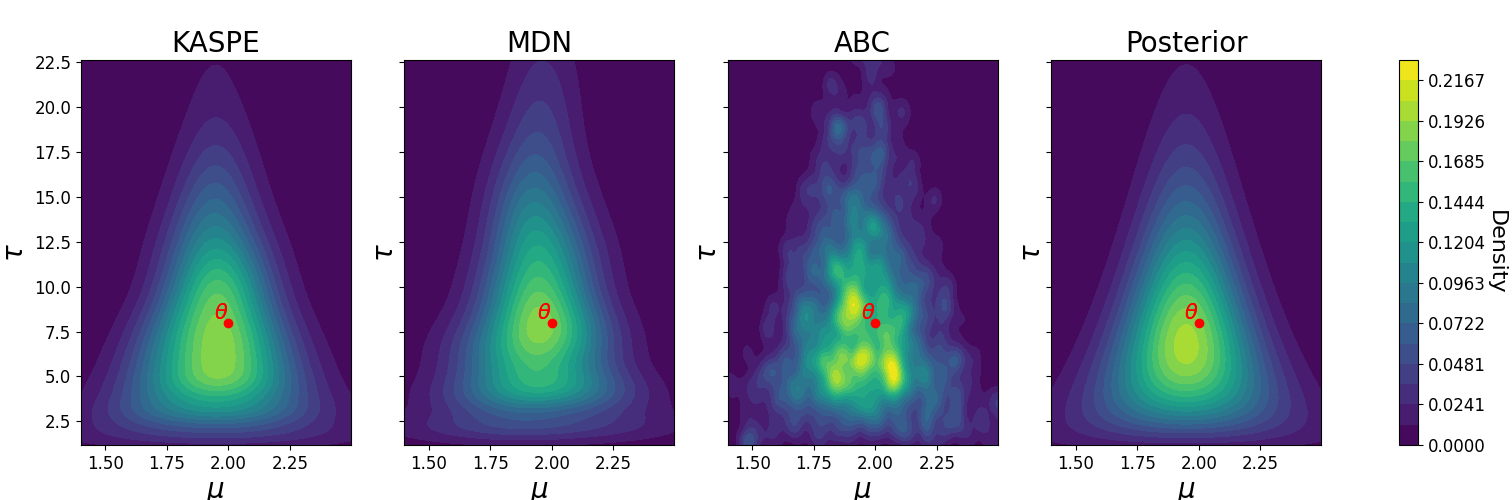}
		}
        \vspace{-0.2in}
	\end{center}
	\caption{Posterior estimation (marginal densities in the first two rows, joint density in last row) of KASPE, MDN, and ABC (columns), respectively, for the Gaussian model with unknown mean and precision example ($m=4$).}
	\label{NG1}
\end{figure}

\begin{figure}[ht]
	\begin{center}
		\subfigure{%
			
			\includegraphics[width=0.7\textwidth]{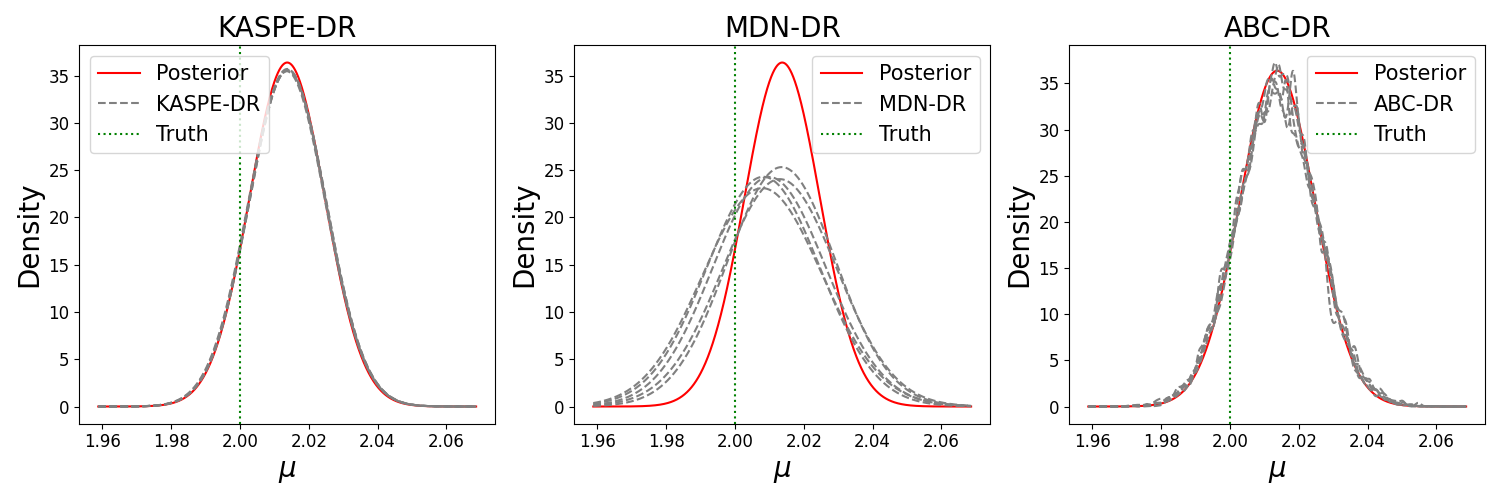}
		}\\
        \vspace{-0.2in}
		\subfigure{%
			
			\includegraphics[width=0.7\textwidth]{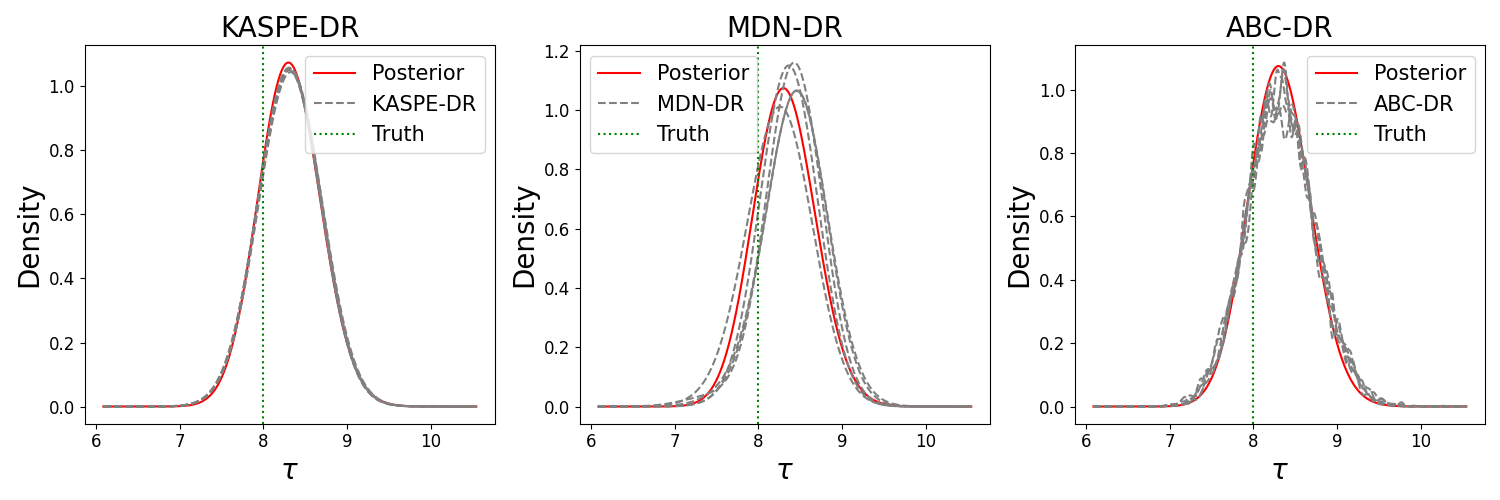}
		}\\
        \vspace{-0.2in}
		\subfigure{%
			
			\includegraphics[width=0.8\textwidth]{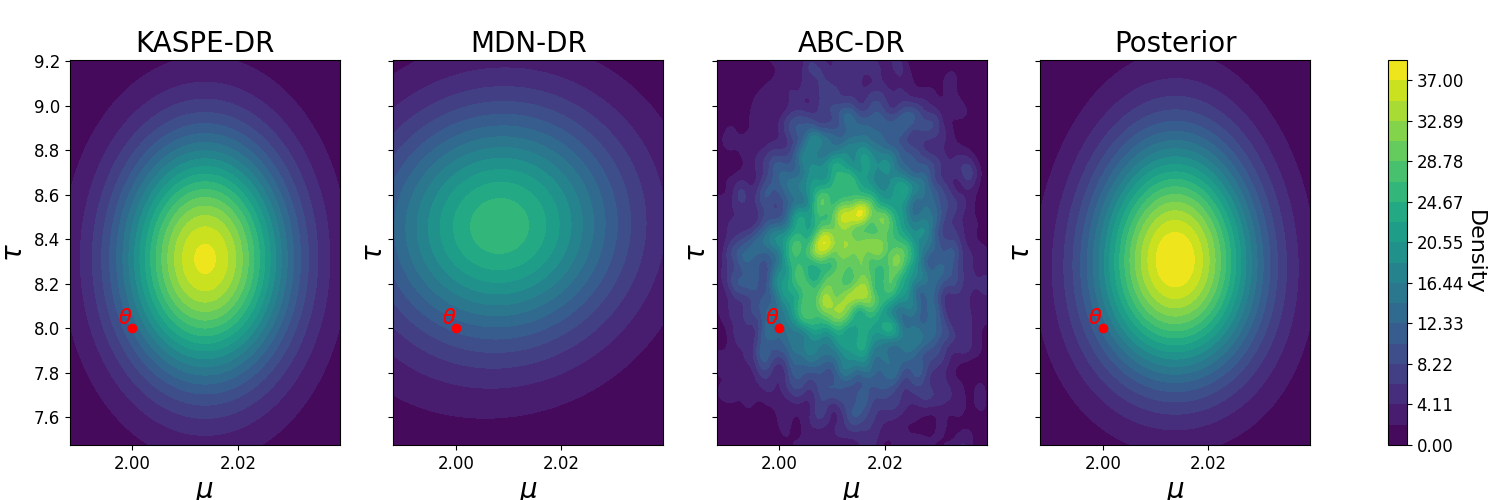}
		}
        \vspace{-0.2in}
	\end{center}
	\caption{Posterior estimation (marginal densities in the first two rows, joint density in last row) of KASPE-DR, MDN-DR, and ABC-DR (columns), respectively, for the Gaussian model with unknown mean and precision example ($m=1,000$).}
	\label{NG4}
\end{figure}

To induce a highly skewed posterior density, we fit a conjugate normal-gamma model with unknown mean $\mu$, precision $\tau$, and a normal-gamma prior to independent normal observations. The simulation experiment reveals that KASPE consistently outperforms the other methods across all scenarios. In particular, Figure \ref{NG1} shows that when $m=4$, the KASPE posterior is nearly indistinguishable from the true posterior density, effectively capturing the heavy-tails of the marginal over $\mu$ and the skewness of the marginal over $\tau$. In contrast, the MDN approach, while generally accurate, exhibits more bias, particularly in regions of high posterior density. This deviation suggests potential limitations in MDN's ability to fully adapt to the posterior landscape. The ABC method captured the overall shape of the posterior distribution but was marked by considerable variability in its density estimate. These characteristics are attributable to the local nature of ABC, which, while effective in capturing broad patterns, can struggle with accuracy and smoothness, particularly in data-sparse regions. As expected, increasing data dimension degrades posterior estimates due to increased Mote Carlo error in ABC, and in the NN training stage of KASPE and MDN since they target a function with increasingly large input space. Nevertheless, when $m=100$, KASPE estimates remain quite accurate (see Supplement Figure S.3), while $m=1,000$ sees estimation performance decrease substantially (see Supplement Figure S.4). Following our recommendation to consider summary statistics rather than the raw data, we implement dimension-reduced versions of the three approaches with two summary statistics: the sample mean and sample variance of the data. Figure \ref{NG4} shows that posterior estimation accuracy substantially improves for all three methods after summarization. KASPE-DR is very accurate, while MDN-DR estimation still suffers from some bias. The pronounced difference in estimation performance between KASPE-DR and MDN-DR can largely be attributed to KASPE-DR’s kernel sampling mechanism, which generates proposals that are more closely aligned with the observed data.  
It is important to note that, as the dimension of observed data increases, the marginal posterior over $\mu$ becomes less heavy-tailed, while the marginal over $\tau$ becomes less skewed, as the relative impact of the prior decreases (see Supplement Sec. S.4.1). Furthermore, for this simple example, the sample mean and sample variance are sufficient for $\mu$ and $\tau$. We note, however, that sufficient statistics will not in general be available in the likelihood-free setting, requiring summaries to be chosen heuristically.

\begin{figure}[ht]
	\begin{center}
		\subfigure{%
			
			\includegraphics[width=0.7\textwidth]{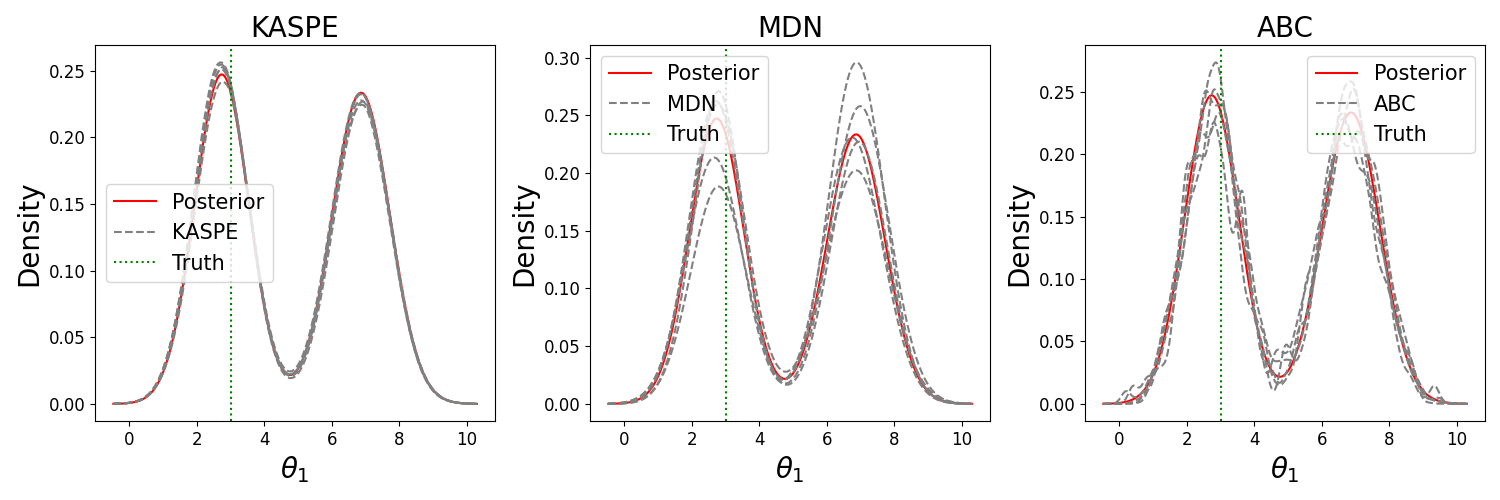}
		}\\
        \vspace{-0.2in}
		\subfigure{%
			
			\includegraphics[width=0.7\textwidth]{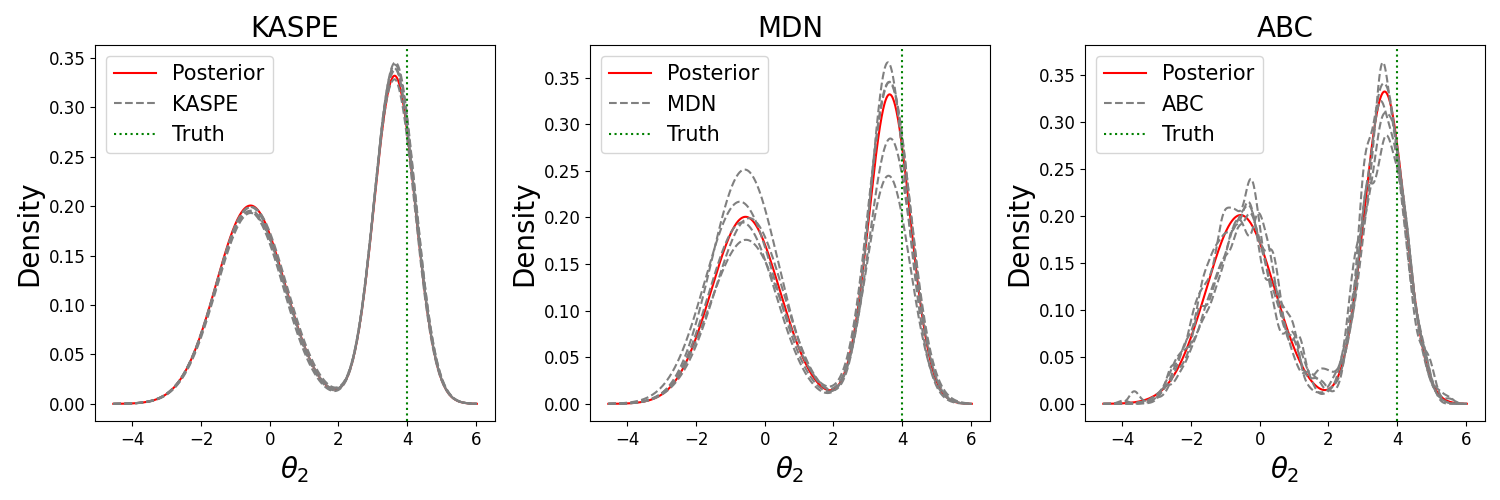}
		}\\
        \vspace{-0.2in}
		\subfigure{%
			
			\includegraphics[width=0.8\textwidth]{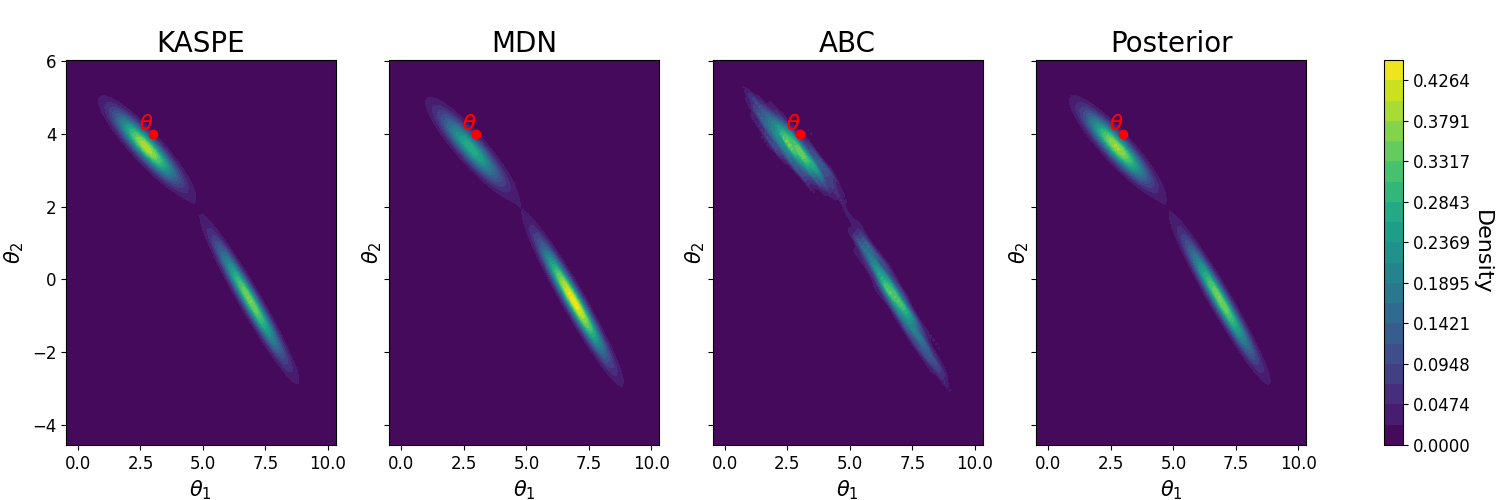}
		}
        \vspace{-0.2in}
	\end{center}
 	\caption{Posterior estimation (marginal densities in the first two rows, joint density in last row) of KASPE, MDN, and ABC (columns), respectively, for the Gaussian mixture model with unknown mean example ($m=4$).}
	\label{Gaussian mixture1}
\end{figure}

Next we consider a model that admits posterior densities with multiple local modes. The observation model is mixture of two Gaussians with fixed mixture weights. The mixture components have fixed covariances and means $v\theta$ and $r\theta$, respectively, where $\theta = (\theta_1, \theta_2)^\top$ is unknown and $v$ and $r$ are fixed covariate vectors. 
Figure \ref{Gaussian mixture1} summarizes posterior estimation performance for all three approaches when \(m=4\). The KASPE posterior is remarkably close to the true posterior, accurately identifying the two modes of the marginal density and capturing the two ellipses in the contour plot of the bivariate density. In contrast, the MDN and ABC posteriors show lower accuracy and greater variability. Similarly to the previous simulation example, posterior estimates under ABC are noticeably less smooth. 

\begin{figure}[ht]
	\begin{center}
		\subfigure{%
			
			\includegraphics[width=0.7\textwidth]{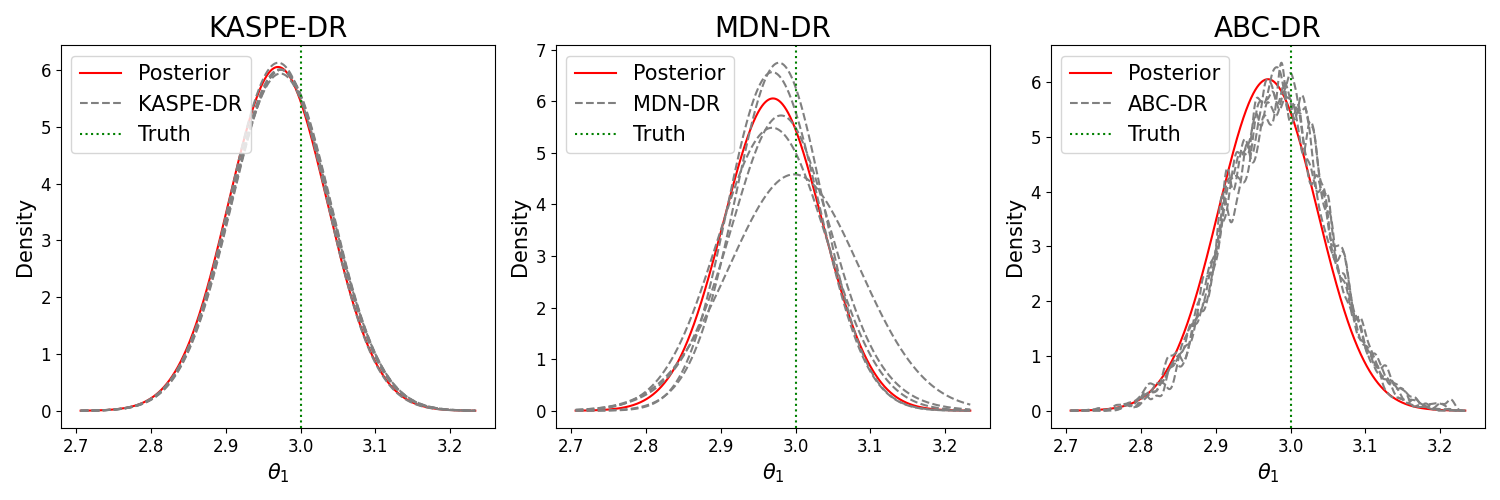}
		}\\
        \vspace{-0.2in}
		\subfigure{%
			
			\includegraphics[width=0.7\textwidth]{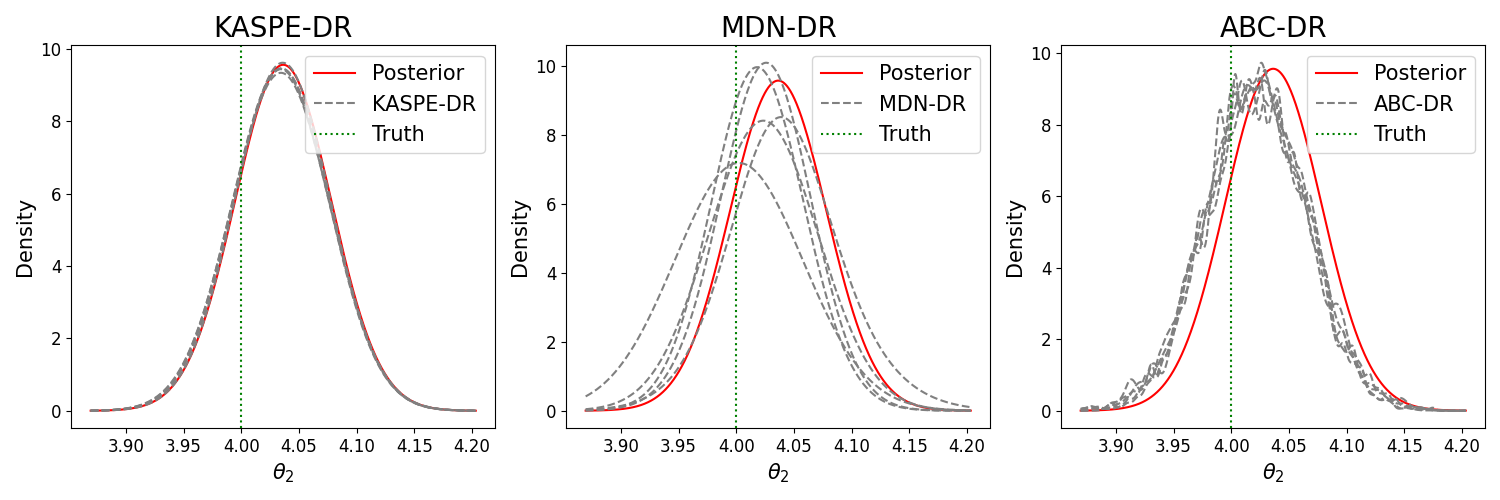}
		}\\
        \vspace{-0.2in}
		\subfigure{%
			
			\includegraphics[width=0.8\textwidth]{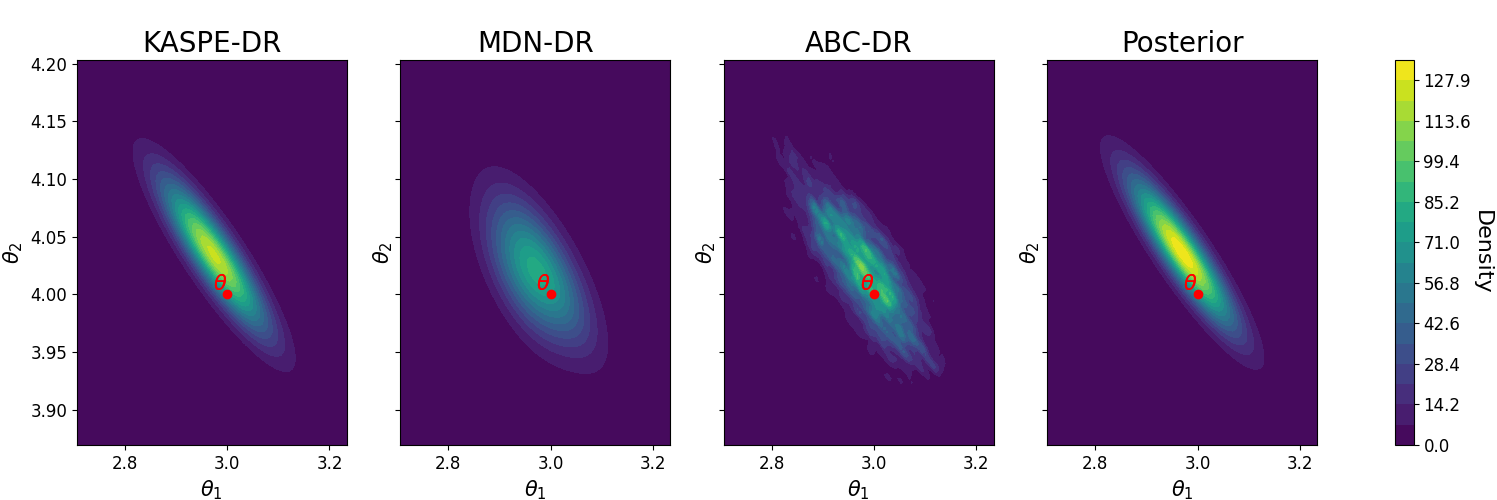}
		}
        \vspace{-0.2in}
	\end{center}
 	\caption{Posterior estimation (marginal densities in the first two rows, joint density in last row) of KASPE-DR, MDN-DR, and ABC-DR (columns), respectively, for the Gaussian mixture model with unknown mean example ($m=1,000$).}
	\label{Gaussian mixture4}
\end{figure}

As the dimension of the data increases to $m = 100$, as shown in Supplement Figure S.6, the number of modes becomes one due to overwhelming location information in the data. Under this scenario, even without dimension reduction, KASPE continues to produce accurate posterior estimates, effectively capturing the bell shape of the posterior marginal density and the elliptical shape of the bivariate density. In contrast, MDN shows a larger deviation from the true posterior, while ABC's estimates are notably poor, demonstrating significant inaccuracies and a failure to capture the shape of the posterior density. Again, as expected, Supplement Figure S.6 shows that when $m=1,000$, the estimation accuracy of all methods deteriorates markedly. KASPE still produces posterior estimates that are closest to the truth, while ABC performs very poorly. To implement the dimension-reduced methods we summarize the data via the least square estimates in the two model components (see Supplement Sec. S.4.2), since the likelihood is a mixture. This results in much better performance for all methods, as illustrated in Figure \ref{Gaussian mixture4} where KASPE-DR can now almost fully recover the truth of the marginal and joint densities, while ABC-DR and MDN-DR have noticeably higher bias and variation.

\begin{figure}[ht]
	\begin{center}
	\includegraphics[width=0.9\textwidth]{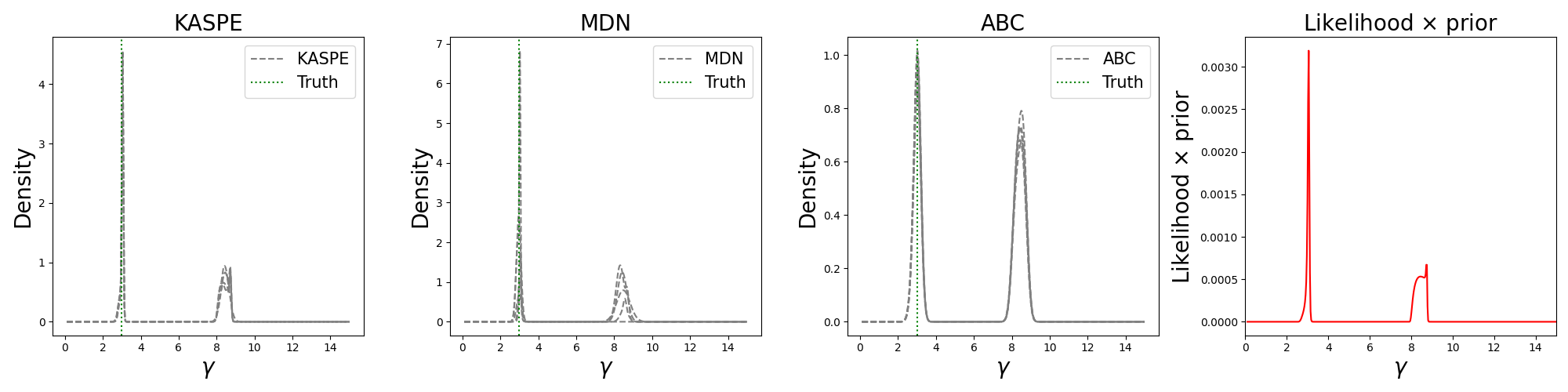}
	\includegraphics[width=0.9\textwidth]{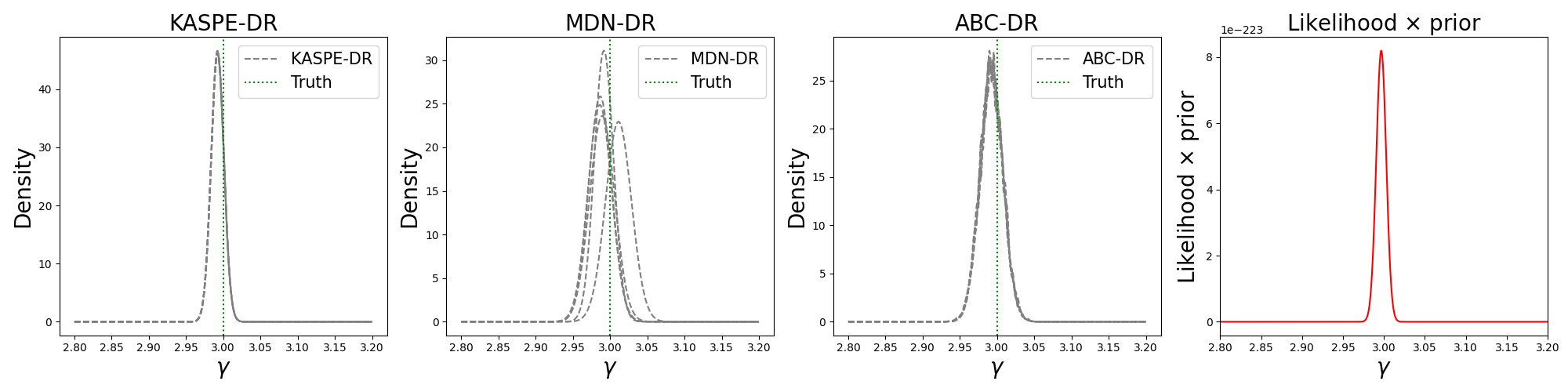}
    \vspace{-0.3in}
	\end{center}
 	\caption{Marginal posterior estimation of (top row, $m=6$) KASPE, MDN, and ABC; and (bottom row, $m=1,000$) KASPE-DR, MDN-DR, and ABC-DR, respectively, for the FN model example.}
	\label{FN4}
\end{figure}

Finally, we consider the problem of fitting a nonlinear dynamical system to noisy data. The FitzHugh-Nagumo model relates membrane voltage $v(t)$ and recovery $r(t)$ in a biological neuron over time $t$ through a system of coupled nonlinear ordinary differential equations. 
All model parameters are assumed fixed, except $\gamma$, which qualitatively contributes to the phase of the limit cycle oscillations in $v$ and $r$. 
Although the likelihood in this example is tractable, the posterior is only known up to a proportion given by the product of the likelihood and prior density. 
The top row of Figure \ref{FN4} shows that when \(m=6\), the posterior exhibits two well-separated modes. KASPE accurately identifies both modes, including their relative magnitudes and spreads. In contrast, MDN shows lower accuracy and higher variability, with one replication failing to capture the second mode altogether. While ABC identifies posterior bi-modality, it produces modes with incorrect relative weight and spread.
As \(m\) increases to 100 (Supplement Figure S.7), the posterior converges to a single mode near the true value of \(\gamma\). Here, KASPE estimation of the posterior density shape and spread is nearly indistinguishable from the true posterior, while MDN estimates remain noticeably less accurate. ABC performs poorly, incorrectly estimating the mode of the density. This degradation in performance, particularly for ABC, can be attributed to the increased dimensionality of the data, which complicates the matching of synthetic and observed data effectively.
Even when $m$ increases to 1,000 (see Supplement Figure S.8), posterior shape and spread are correctly identified by KASPE, while MDN inflates the range of non-negligible positive posterior density, and ABC cannot even roughly capture the region with positive posterior density. It is still worthwhile to evaluate the impact of dimension reduction in this example. Since the ODE solution has a periodic pattern, we summarize the data by the estimated coefficients of a Fourier basis expansion with 11 basis functions. 
Results in the bottom row of Figure \ref{FN4} show that KASPE estimates remain precise, while MDN-DR is more variable and inflates posterior spread. ABC-DR benefits substantially from dimension reduction, becoming much less variable and smoother. More importantly, it is able to capture the region of high posterior density, although it remains less accurate than KASPE-DR.

\section{Summary}\label{ch3.discussion}
This work introduces the KASPE technique for uncertainty quantification in inference problems, which recasts inference as an optimization task, providing an alternative to Monte Carlo sampling-based ABC methods within the framework of fully Bayesian statistical inference. KASPE leverages a deep neural network to map data to the parameters of a specified family of distributions. This is complemented by a kernel-based adaptive sampling mechanism that selectively accepts synthetic training data based on their similarity to observed data, thereby refining the synthetic dataset and enhancing posterior density estimation. KASPE is straightforward to implement, broadly applicable even when the likelihood function is computationally intractable, and produces a direct closed-form approximation of the posterior density. We study its large sample properties and connections with other existing likelihood-free and likelihood-based approaches. Simulation experiments indicate that KASPE consistently outperforms competing methods, effectively capturing complex posterior landscapes, including multi-modality and heavy-tails. While all methods may experience performance degradation with increasing data dimension, KASPE mitigates this via dimension reduction, improving both the accuracy and computational efficiency of posterior estimation. 

For high-dimensional data where informative application-specific summaries may not be readily available, a promising research direction is to automate learning of important features or summary statistics. This could be achieved through unsupervised learning techniques, such as autoencoders, to reduce the dimensionality of the data before inputting it into the neural network model of KASPE. Alternatively, one may incorporate a transformer encoder into our neural network model directly to summarize key features for sequential data, leveraging its ability to capture long-range dependencies and contextual relationships through self-attention mechanisms, which may enhance the model's capacity to learn rich representations and improve the robustness and accuracy of posterior density estimation.

\bibliographystyle{agsm}
\bibliography{ref}

\end{document}


\def\spacingset#1{\renewcommand{\baselinestretch}%
{#1}\small\normalsize} \spacingset{1}


\if1\blind
{
  \title{\bf Supplement for "Likelihood-free Posterior Density Learning for Uncertainty Quantification in Inference Problems"}
  \author{Rui Zhang 
    Department of Statistics, The Ohio State University\\
    and \\
    Oksana A Chkrebtii \\
    Department of Statistics, The Ohio State University}
  \maketitle
} \fi

\if0\blind
{
  \bigskip
  \bigskip
  \bigskip
  \begin{center}
    {\LARGE\bf Supplement for "Likelihood-free Posterior Density Learning for Uncertainty Quantification in Inference Problems"}
\end{center}
  \medskip
} \fi

\newpage
\spacingset{1.9} 

\section{Additional Figures}

\subsection{Graphical Representation of NN Architecture}
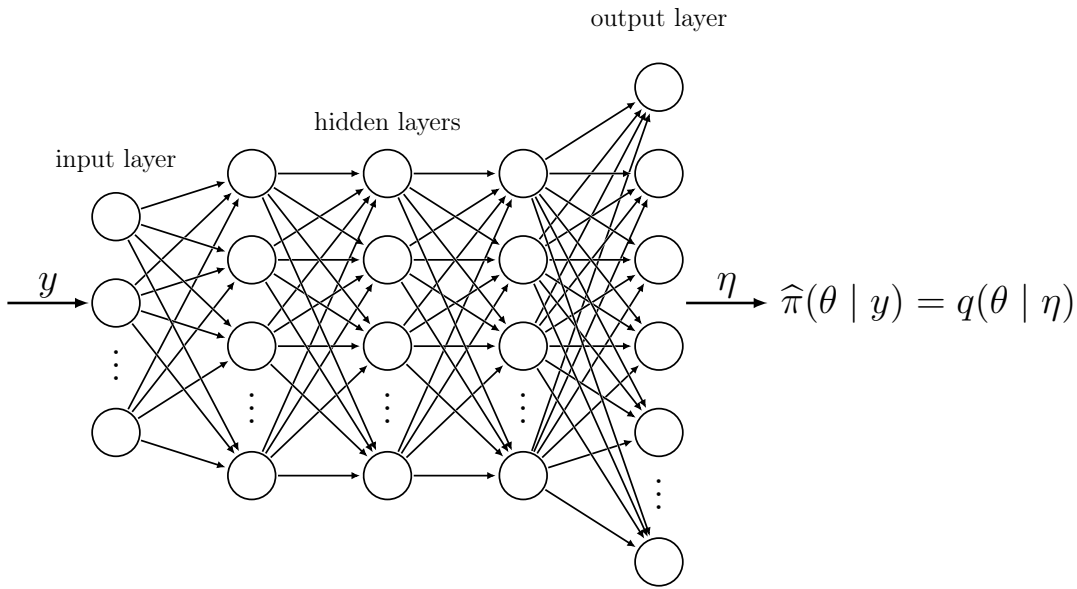
\begin{figure}[ht!]
\resizebox{0.9\textwidth}{!}{ 
\begin{tikzpicture}[x=2.2cm,y=1.4cm] 
\def\yshift{0.5} 
   \node[scale=1.5] at (0.5,-\yshift+0.2) {$y$};
  \draw[->,black,line width=1.2] (0.2,-\yshift) -- (0.8,-\yshift);
  \message{^^JNeural network, shifted}
  \readlist\Nnod{3,4,4,4,6} 
  \readlist\Nstr{n,m,m,m,\nu} 
  \readlist\Cstr{\strut x,a^{(\prev)},a^{(\prev)},a^{(\prev)},\eta} 
  \def\lastlayershift{0.8} 
  
  \message{^^J  Layer}
  \foreachitem \N \in \Nnod{ 
    \def\lay{\Ncnt} 
    \pgfmathsetmacro\prev{int(\Ncnt-1)} 
    \message{\lay,}
    \foreach \i [evaluate={\c=int(\i==\N); \d=int(\i>3)+int(\i>6);
                 \btwnodeshift=(\lay==\Nnodlen ? \lastlayershift : 1);
                 \y=\N/2-\i-\c*\yshift;
                 \index=(\i<\N?int(\i):"\Nstr[\lay]");
                 \x=\lay; \n=\nstyle;}] in {1,...,\N}{ 
      \node[node \n] (N\lay-\i) at (\x,\y) {};
      
      \ifnum\lay>1 
        \foreach \j in {1,...,\Nnod[\prev]}{ 
          \draw[connect,white,line width=1.2] (N\prev-\j) -- (N\lay-\i);
          \draw[connect arrow] (N\prev-\j) -- (N\lay-\i);
        }
      \fi 

    }
    \path (N\lay-\N) --++ (0,1+\yshift) node[midway,scale=1.5] {$\vdots$};
}
  \node[above=5,align=center,black] at (N1-1.90) {input layer};
  \node[above=2,align=center,black] at (N3-1.90) {hidden layers};
  \node[above=10,align=center,black] at (N\Nnodlen-1.90) {output layer};
  \draw[->,black,line width=1.2] (\Nnodlen+0.2,-\yshift) -- (\Nnodlen+0.8,-\yshift);
  \node[align=center,anchor=west,scale=1.5] at (\Nnodlen+0.8,-\yshift) {$\hat{\pi}
  (\theta \mid y)=q(\theta \mid \eta)$};
  \node[scale=1.5] at (\Nnodlen+0.5,-\yshift+0.2) {$\eta$};
\end{tikzpicture}
}
	\caption{Graphical representation of the KASPE model: The feed-forward neural network takes data 
$y$ as input, and its output determines parameters $\eta$ for a family of densities. The density serves as an estimate of the posterior density conditional on the input data.}
	\label{fig:KASPE structure}
\end{figure}

Figure \ref{fig:KASPE structure} illustrates the NN architecture for the KASPE method.

\section{Proofs}

\subsection{Proof of Theorem 1}

\begin{proof}
Let \((\Xi, \mathcal{F}, P)\) denote the underlying probability space with respect to which all convergence notions are defined, where \(\Xi\) is the sample space (set of all outcomes), \(\mathcal{F}\) is a $\sigma$-algebra of measurable subsets of \(\Xi\), and \(P\) is a probability measure on \(\mathcal{F}\).

To prove convergence in probability, it suffices to show that any subsequence of $\{\hat{\omega}_n\}$ has a further subsequence along which the corresponding functions converge pointwise almost surely to $\mathbf{N}_0(\cdot)$. This implies convergence pointwise in probability of the entire sequence $\mathbf{N}(\cdot, \hat{\omega}_n)$ to $\mathbf{N}_0(\cdot)$ as $n\to\infty$.

Let \( \{\hat{\omega}_{n_k}\} \) be an arbitrary subsequence of \( \{\hat{\omega}_n\} \). Since each \( \hat{\omega}_{n_k} \) takes values in the compact set \( \Omega \), by compactness there exists a further subsequence \( \{\hat{\omega}_{m_j}\} \) and a random variable \( \omega^* \) such that:
\[
\hat{\omega}_{m_j} \xrightarrow{a.s.} \omega^*, \text{ as } j\to \infty.
\]
Define $\Xi_1 := \left\{ \xi \in \Xi : \lim_{j \to \infty}\hat{\omega}_{m_j}(\xi) = \omega^*(\xi) \right\},$
so that \( P(\Xi_1) = 1 \). Next, by the uniform convergence assumption,
\[
\sup_{\omega \in \Omega} |Q_n(\omega) - Q_0(\omega)| \xrightarrow{p} 0 , \text{ as } n\to \infty, 
\]
and by the subsequence principle, there exists a further subsequence of \( \{m_j\} \) (which we continue to denote by \( \{m_j\} \) for notational simplicity) such that:
\[
\sup_{\omega \in \Omega} |Q_{m_j}(\omega) - Q_0(\omega)| \xrightarrow{a.s.} 0, \text{ as } j\to \infty.
\]
Denote \( Q_n(\omega, \xi) \) as the realization of the random function \( Q_n(\omega) \) at the outcome \( \xi \in \Xi \). Define $
\Xi_2 := \left\{ \xi \in \Xi :  \lim_{j \to \infty}\sup_{\omega \in \Omega} |Q_{m_j}(\omega, \xi) - Q_0(\omega)| = 0 \right\},$ so that $ P(\Xi_2) = 1$. And we let 
$\Xi' := \Xi_1 \cap \Xi_2,$ so that $P(\Xi') = 1.$

Now, for any $\xi \in \Xi'$, by definition of $\hat{\omega}_{m_j}(\xi)$ as a minimizer,
$Q_{m_j}(\hat{\omega}_{m_j}(\xi), \xi) \leq Q_{m_j}(\omega_0, \xi)$ for any $\omega_0 \in \Omega_0$. Since \(\lim_{j \to \infty}\sup_{\omega \in \Omega} |Q_{m_j}(\omega, \xi) - Q_0(\omega)| =0\) for all \(\xi \in \Xi'\), it follows that for any fixed \(\omega \in \Omega\), \(\lim_{j \to \infty}Q_{m_j}(\omega, \xi) = Q_0(\omega)\). Moreover, since \(\lim_{j \to \infty}\hat{\omega}_{m_j}(\xi) =\omega^*(\xi)\), and the convergence of \(Q_{m_j}(\omega, \xi)\) to \(Q_0(\omega)\) is uniform in \(\omega\), we can conclude that \(\lim_{j \to \infty}Q_{m_j}(\hat{\omega}_{m_j}(\xi), \xi) = Q_0(\omega^*(\xi))\). On the other hand, since \(\omega_0\) is fixed, we have \(\lim_{j \to \infty}Q_{m_j}(\omega_0, \xi) = Q_0(\omega_0)\). Because the inequality \(Q_{m_j}(\hat{\omega}_{m_j}(\xi), \xi) \leq Q_{m_j}(\omega_0, \xi)\) holds for every \(j\) and both sides converge,  we pass to the limit in the inequality, yielding \(Q_0(\omega^*(\xi)) \leq Q_0(\omega_0)\) for all \(\xi \in \Xi'\).

Since $\omega_0$ is a minimizer of $Q_0$, the inequality implies:
    \[
    Q_0(\omega^*(\xi)) = Q_0(\omega_0),
    \]
    showing that $\omega^*(\xi) \in \Omega_0$ for all $\xi \in \Xi'$. By the unique minimizing NN function assumption, any $\omega \in \Omega_0$ induces the same function
$\mathbf{N}(\cdot, \omega) = \mathbf{N}_0(\cdot). $
Therefore, we conclude that
\begin{equation}\label{NN minimization function}
    \mathbf{N}(\cdot, \omega^*(\xi)) = \mathbf{N}_0(\cdot)
\end{equation}
for all \(\xi \in \Xi'\), which holds almost surely. Finally, by the continuity of \(\mathbf{N}(\cdot, \omega)\) in \(\omega\) and the almost sure convergence \(\hat{\omega}_{m_j} \xrightarrow{a.s.} \omega^*\), we have that for each fixed $y\in \mathcal{Y}$, \(\mathbf{N}(y, \hat{\omega}_{m_j}) \xrightarrow{a.s.} \mathbf{N}(y, \omega^*)\) as $j \to \infty$. Since we have established that \(\mathbf{N}(\cdot, \omega^*) = \mathbf{N}_0(\cdot)\) almost surely, we conclude that for each fixed $y\in \mathcal{Y}$,
\[
\mathbf{N}(y, \hat{\omega}_{m_j}) \xrightarrow{a.s.} \mathbf{N}_0(y), \text{ as } j \to \infty.
\]

Since the original subsequence \(\{\hat{\omega}_{n_k}\}\) was arbitrary, we have shown that every subsequence of \(\{\hat{\omega}_n\}\) admits a further subsequence along which the corresponding functions converge pointwise almost surely to \(\mathbf{N}_0(\cdot)\). This implies that the entire sequence \(\mathbf{N}(\cdot, \hat{\omega}_n)\) converges pointwise in probability to \(\mathbf{N}_0(\cdot)\) as $n\to\infty$, that is, for each fixed $y\in \mathcal{Y}$,
\[
\hat{\eta}_n(y) \xrightarrow{p} \mathbf{N}_0(y), \text{ as } n\to\infty.
\]
This completes the proof.
\end{proof}

\subsection{Proof of Theorem 2}
\begin{proof}
By the definition of the expected training loss function,
$$
\begin{aligned}
Q_0(\omega)   &= -\mathbb{E}_{(\theta, y, w)}\bigl[w\log q(\theta \mid \mathbf{N}(y,\omega))\bigr]\\
&=-\mathbb{E}_{(\theta, y)}\bigl[K(\frac{y-y_0}{h})\log q(\theta \mid \mathbf{N}(y,\omega))\bigr] \\
&=\mathbb{E}_{(\theta, y)}\bigl[K(\frac{y-y_0}{h})\log \frac{\pi(\theta \mid y)}{q(\theta \mid \mathbf{N}(y,\omega))}\bigr]-\mathbb{E}_{(\theta, y)}\bigl[K(\frac{y-y_0}{h})\log\pi(\theta \mid y)\bigr]\\
&=\mathbb{E}_{y}\bigl[K(\frac{y-y_0}{h})\mathbb{E}_{\theta \mid y}\bigl[\log \frac{\pi(\theta \mid y)}{q(\theta \mid \mathbf{N}(y,\omega))}\bigr]\bigr]-\mathbb{E}_{\theta, y}\bigl[K(\frac{y-y_0}{h})\log\pi(\theta \mid y)\bigr]\\
&=\mathbb{E}_{y}\bigl[K(\frac{y-y_0}{h})\text{KL}\left( \pi(\cdot \mid y)\ \mid  q(\cdot \mid \mathbf{N}(y,\omega)\right)\bigr]-\mathbb{E}_{\theta, y}\bigl[K(\frac{y-y_0}{h})\log\pi(\theta \mid y)\bigr].
\end{aligned}
$$

Since the second term is constant with respect to $\omega$, minimizing $Q_0(\omega)$ is equivalent to minimizing $\mathbb{E}_{y}\bigl[K(\frac{y-y_0}{h})\text{KL}\left( \pi(\cdot \mid y)\ \mid  q(\cdot \mid \mathbf{N}(y,\omega)\right)\bigr]$ with respect to $\omega$. By Assumption 3 in Theorem 1, there exists $\mathbf{N}_0(\cdot)\in \mathcal{M}$ such that for any $\omega \in \Omega$, 
$$\mathbb{E}_{y}\bigl[K(\frac{y-y_0}{h})\text{KL}\left( \pi(\cdot \mid y)\ \mid  q(\cdot \mid \mathbf{N}_0(y)\right)\bigr]\leq \mathbb{E}_{y}\bigl[K(\frac{y-y_0}{h})\text{KL}\left( \pi(\cdot \mid y)\ \mid  q(\cdot \mid \mathbf{N}(y,\omega)\right)\bigr].$$
Since $\eta(\cdot)\in \mathcal{M}$, it follows that 
\begin{equation}\label{inequality1}
    \mathbb{E}_{y}\bigl[K(\frac{y-y_0}{h})\text{KL}\left( \pi(\cdot \mid y)\ \mid  q(\cdot \mid \mathbf{N}_0(y)\right)\bigr]\leq \mathbb{E}_{y}\bigl[K(\frac{y-y_0}{h})\text{KL}\left( \pi(\cdot \mid y)\ \mid  q(\cdot \mid \eta(y)\right)\bigr].
\end{equation}

By the definition of the EP parameterization estimator, for any $\omega \in \Omega$,
$$\text{KL}\left(\pi(\cdot \mid y) \mid q(\cdot \mid \eta(y))\right) \leq\text{KL}\left(\pi(\cdot \mid y) \mid q(\cdot \mid \mathbf{N}(y,\omega))\right),$$
and since the kernel function is non-negative, we have
$$\mathbb{E}_{y}\bigl[K(\frac{y-y_0}{h})\text{KL}\left(\pi(\cdot \mid y) \mid q(\cdot \mid \eta(y))\right)\leq \mathbb{E}_{y}\bigl[K(\frac{y-y_0}{h})\text{KL}\left( \pi(\cdot \mid y)\ \mid  q(\cdot \mid \mathbf{N}(y,\omega)\right)\bigr]$$
for any $\omega \in \Omega$. In particular, since $\mathbf{N}_0(\cdot)\in \mathcal{M}$, it follows that
\begin{equation}\label{inequality2}
\mathbb{E}_{y}\bigl[K(\frac{y-y_0}{h})\text{KL}\left(\pi(\cdot \mid y) \mid q(\cdot \mid \eta(y))\right)\leq \mathbb{E}_{y}\bigl[K(\frac{y-y_0}{h})\text{KL}\left( \pi(\cdot \mid y)\ \mid  q(\cdot \mid \mathbf{N}_0(y)\right)\bigr].
\end{equation}
Combining inequalities (\ref{inequality1}) and (\ref{inequality2}) and using the uniqueness of $\mathbf{N}_0(\cdot)$ under Assumption 3, we conclude that $\mathbf{N}_0(\cdot) =\eta(\cdot)$. By Theorem 1, for each fixed $y\in \mathcal{Y}$, we have $\hat{\eta}_n(y)\xrightarrow{p} \mathbf{N}_0(y), \text{ as }n \to \infty.$ Thus, $\hat{\eta}_n(y)\xrightarrow{p} \eta(y), \text{ as }n \to \infty.$
This completes the proof.
\end{proof}

\section{Additional Settings for Simulation Experiments}

Candidate families of densities for KASPE and MDN consist of a mixture of 20 Gaussian densities with full covariance matrices. For the NN architecture, we use a feed-forward network with 2 hidden layers. The training sample size is $n = 125,000$, of which $25\%$ is held out for validation. For KASPE and ABC, a squared exponential kernel is used to measure distance between observed and synthetic data, with the bandwidth parameter chosen manually to be as small as possible while targeting the desired acceptance rate. ABC is implemented using adaptive tuning of the proposal covariance within a parallel-tempering ABC-MCMC algorithm to enable the sampler to efficiently explore posteriors with possible local modes \citep{Swendsen1986, Geyer1991}. Convergence is assessed by monitoring traceplots and correlation plots. The ABC posterior density is approximated from the ABC sample via kernel density estimation. All methods were simulated 5 times for each scenario, each time with the same data.

\subsection{Skewed Posterior Density with Heavy Tails}
Consider i.i.d. data $y = (y(1), \ldots, y(m))^\top$ from $\mathcal{N}(\mu, \tau^{-1})$, where the mean, $\mu$, and the precision, $\tau$, follow a conjugate normal-gamma prior. That is, the conditional distribution of $\mu$ given $\tau$ is $\mu  \mid  \tau \sim \mathcal{N}(\eta, (\lambda \tau)^{-1})$, where $\eta$ and $\lambda$ are the prior mean and precision of $\mu$, respectively. The marginal distribution of $\tau$ follows a gamma distribution, $\tau \sim \text{Gamma}(\alpha, \beta)$, with shape and rate parameters $\alpha$ and $\beta$, respectively. We can thus write,
\begin{equation}
    (\mu,\tau)\sim \text{NG}(\eta,\lambda,\alpha,\beta).
\end{equation}
Note that the marginal prior over $\mu$ is a non-standard Student's t-distribution with degrees of freedom $2\alpha$, location parameter $\eta$, and scale parameter $\frac{\beta}{\lambda \alpha}$. 
The posterior over $(\mu,\tau)$ also follows a normal-gamma distribution given by
\begin{equation*}
   \pi(\mu,\tau \mid y) \sim \text{NG}\left(\frac{\lambda\eta+m\bar{y}}{\lambda+m},\lambda+m,\alpha+\frac{m}{2},\; \beta+\frac{1}{2}(ms+\frac{\lambda m (\bar{y}-\eta)^2}{\lambda+m})\right),
\end{equation*}
where $\bar{y} = \frac{1}{m}\sum_{i=1}^{m} y(i)$ is the sample mean and $s = \frac{1}{m}\sum_{i=1}^{m}(y(i) - \bar{y})^2$ is the sample variance. For our numerical experiment, we choose hyperparameter values $\eta=2, \lambda=\frac{1}{16},\alpha=1.01, \beta=0.1$.

\subsection{Posterior Density with Multiple Local Modes}
Consider a Gaussian mixture model with unknown mean components $\theta = (\theta_1, \theta_2)^\top$ and define the time-dependent covariate vectors \( v(t)=(t, t^2)^\top \) and \( r(t)=(t^2, \sqrt{t})^\top \). These vectors are evaluated at time points \( t = (t_1, \ldots, t_m)^\top \), forming the $m\times 2$ design matrices,
\begin{align*}
v &= (v(t_1), v(t_2), \ldots, v(t_m))^\top,\\
r &= (r(t_1), r(t_2), \ldots, r(t_m))^\top.
\end{align*}
The data is a time series \( y = (y(t_1), y(t_2), \ldots, y(t_m))^\top \in \mathbb{R}^m \) and the likelihood for \( y \) given \( \theta \) follows a Gaussian mixture distribution:
\begin{equation}
    y  \mid  \theta \sim p_1 \mathcal{N}(v \theta, \Sigma_1) + p_2 \mathcal{N}(r \theta, \Sigma_2), \label{Gaussian_mixture_model}
\end{equation}
where \( p_1 \) and \( p_2 \) are the mixture probabilities, $v \theta$ and $r \theta$  are 
mean vectors of each component, and \( \Sigma_1 \) and \( \Sigma_2 \) are the associated error covariance matrices. The unknown parameter is \( \theta = (\theta_1, \theta_2)^\top \), while other parameters are fixed: $\Sigma_1=0.3^2 I_{m\times m}, \Sigma_2=0.4^2 I_{m\times m}, p_1=0.4, p_2=0.6$. 
We assume a normal prior for $\theta$ with mean $\mu_0=(2,1)^\top$ and covariance $\Sigma_0=4^2I_{2\times 2}$. The posterior distribution of \( \theta \) can be derived in closed form as
\begin{equation}
  \theta  \mid  y \sim p_1^* \mathcal{N}(\mu_1^*, \Sigma_1^*) + p_2^* \mathcal{N}(\mu_2^*, \Sigma_2^*),
\end{equation}
where,
\[
\Sigma_1^* = (\Sigma_0^{-1} + v' \Sigma_1^{-1} v)^{-1}, \quad \mu_1^* = \Sigma_1^* (\Sigma_0^{-1} \mu_0 + v' \Sigma_1^{-1} y)
\]
\[
\Sigma_2^* = (\Sigma_0^{-1} + r' \Sigma_2^{-1} r)^{-1}, \quad \mu_2^* = \Sigma_2^* (\Sigma_0^{-1} \mu_0 + r' \Sigma_2^{-1} y)
\]
\[
\alpha_1 = p_1 |\Sigma_1|^{-\frac{1}{2}} |\Sigma_1^*|^{\frac{1}{2}} \exp\left(-\frac{1}{2} (y' \Sigma_1^{-1} y - \mu_1^{*'} \Sigma_1^{*-1} \mu_1^*)\right)
\]
\[
\alpha_2 = p_2|\Sigma_2|^{-\frac{1}{2}} |\Sigma_2^*|^{\frac{1}{2}} \exp\left(-\frac{1}{2} (y' \Sigma_2^{-1} y - \mu_2^{*'} \Sigma_2^{*-1} \mu_2^*)\right)
\]
\[
p_1^* = \frac{\alpha_1}{\alpha_1 + \alpha_2}, \quad p_2^* = \frac{\alpha_2}{\alpha_1 + \alpha_2}.
\]

For data reduction, we use least square estimates in each component as summary statistics, namely, $s^\top=((v^\top  v)^{-1}v^\top y,(r^\top r)^{-1}r^\top y)$, since the liklihood is a mixture of two linear models.

\subsection{Posterior Inference for a Dynamical System}
Finally, we consider posterior inference on a parameter of the FitzHugh–Nagumo (FN) model, which is a nonlinear system of ordinary differential equations. This model describes the dynamically spiking membrane potential of a biological neuron. The membrane voltage $v(t)$ and the recovery $r(t)$ evolve over time according to 
\begin{equation}
  \left\{
    \begin{aligned}
       \mathrm{d}v/\mathrm{d} t &=\gamma\left(v-v^{3}/3+r+\zeta\right), \\
       \mathrm{d} r/\mathrm{d} t &=-\gamma^{-1}\left(v-\theta_{1}+\theta_{2} r\right), 
    \end{aligned}
  \right. \label{FN model} 
\end{equation}
starting from initial conditions $v(0)=-1$, and $r(0)=1$. Suppose that we observe data $y=(y(t_1),\ldots,y(t_{m}))^\top\in \mathbb{R}^{m}$, where $y(t)$ is composed of the first component of the ODE solution plus an error term $y(t)=v(t)+\epsilon$ with $\epsilon \stackrel{\text{i.i.d.}}{\sim}\mathcal{N}(0,0.5^2)$. In this example, $\gamma$ is the unknown parameter, while $\theta_{1},\theta_{2} \text{ and }\zeta$ are fixed constants set to  $\theta_{1}=0.2,\theta_{2}=0.2,\zeta=-0.4$. We assume the uniform prior $\gamma\sim \text{unif} (0,15)$.

\section{Additional Figures}

\begin{figure}[ht!]
	\begin{center}
		\subfigure{%
			
			\includegraphics[width=0.8\textwidth]{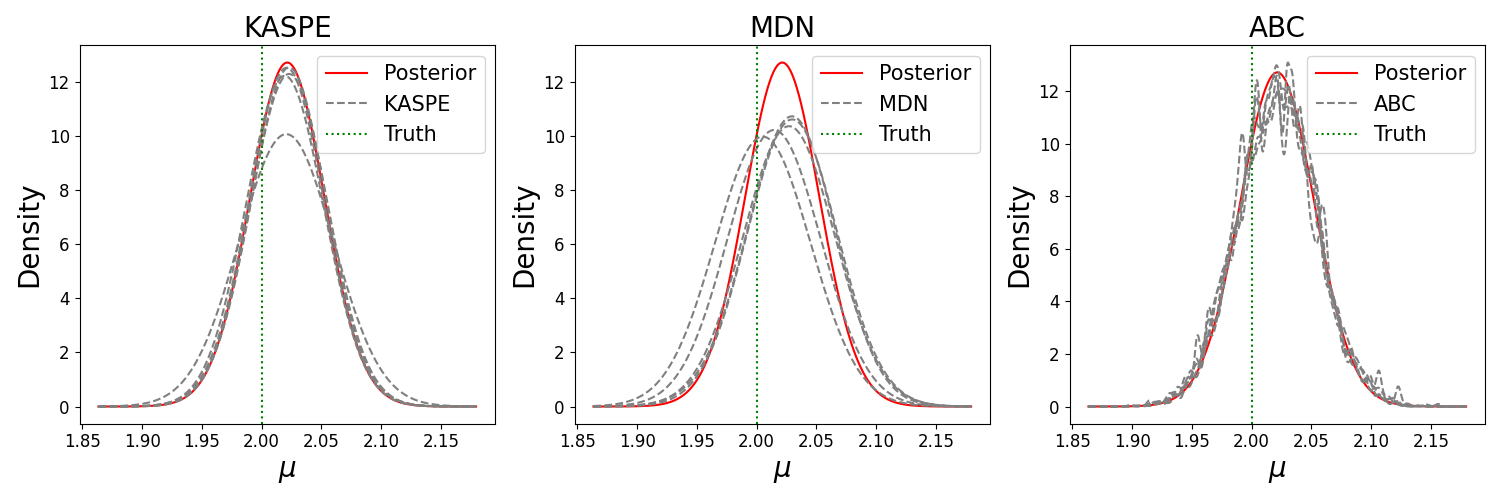}
		}\\
        \vspace{-0.2in}
		\subfigure{%
			
			\includegraphics[width=0.8\textwidth]{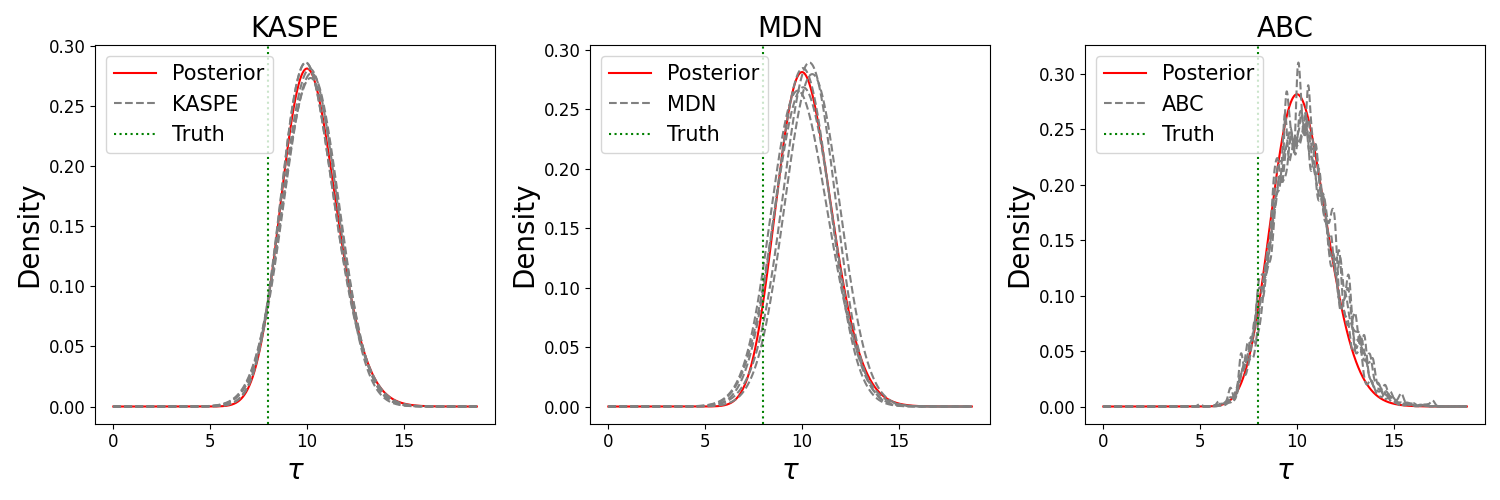}
		}\\
        \vspace{-0.2in}
		\subfigure{%
			
			\includegraphics[width=0.8\textwidth]{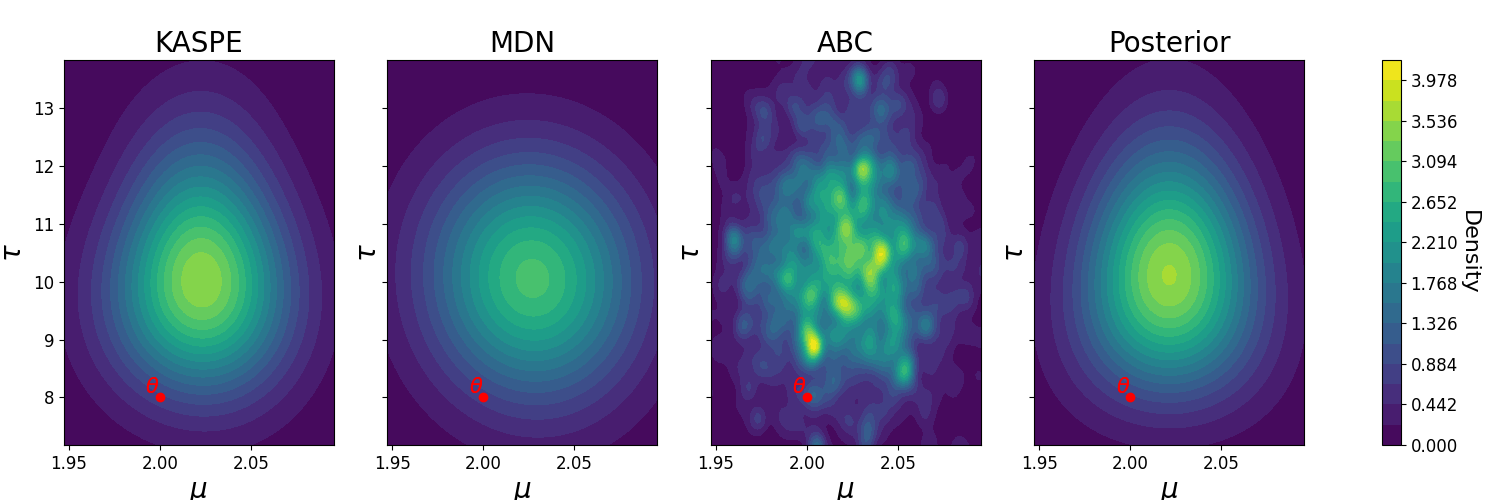}
		}
        \vspace{-0.2in}
	\end{center}
		\caption{Posterior estimation (marginal densities in the first two rows, joint density in last row) of KASPE, MDN, and ABC (columns), respectively, for the Gaussian model with unknown mean and precision example ($m=100$).}
	\label{NG2}
\end{figure}

\begin{figure}[ht!]
	\begin{center}
		\subfigure{%
			
			\includegraphics[width=0.8\textwidth]{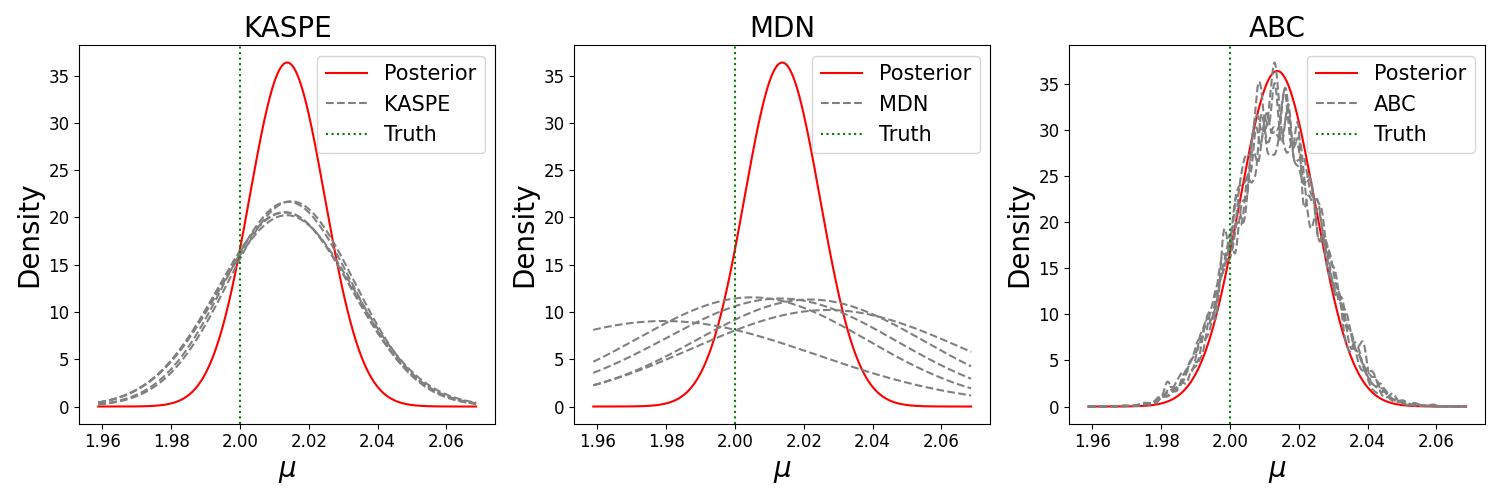}
		}\\
        \vspace{-0.2in}
		\subfigure{%
			
			\includegraphics[width=0.8\textwidth]{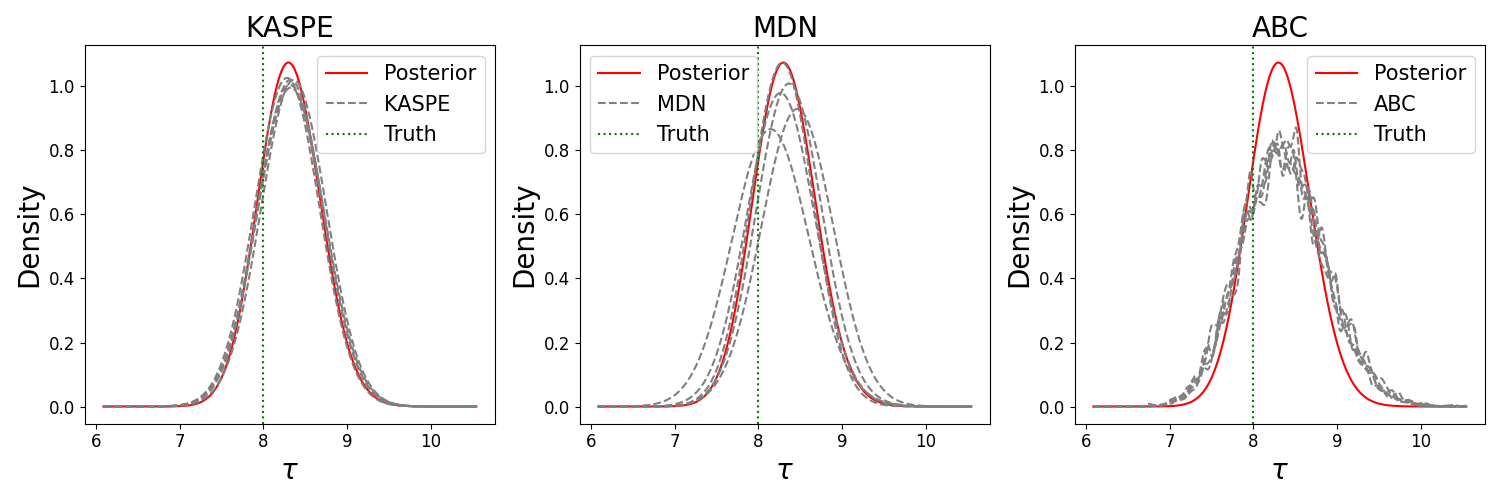}
		}\\
        \vspace{-0.2in}
		\subfigure{%
			
			\includegraphics[width=0.8\textwidth]{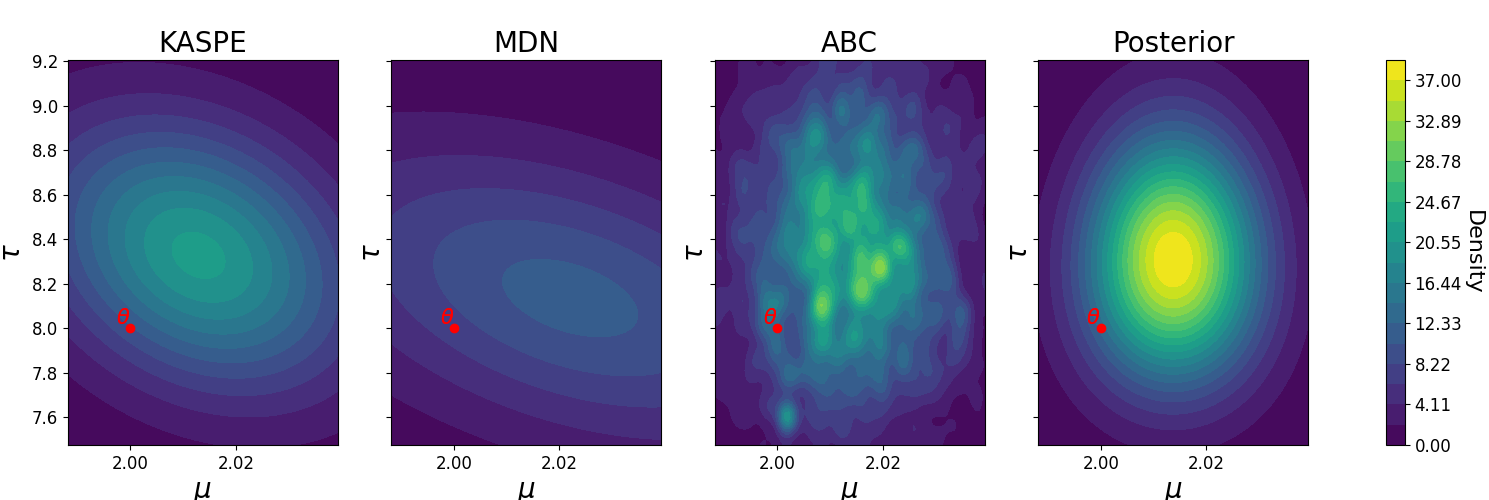}
		}
        \vspace{-0.2in}
	\end{center}
	\caption{Posterior estimation (marginal densities in the first two rows, joint density in last row) of KASPE, MDN, and ABC (columns), respectively, for the Gaussian model with unknown mean and precision example ($m=1,000$).}
	\label{NG3}
\end{figure}


\begin{figure}[ht!]
	\begin{center}
		\subfigure{%
			
			\includegraphics[width=0.8\textwidth]{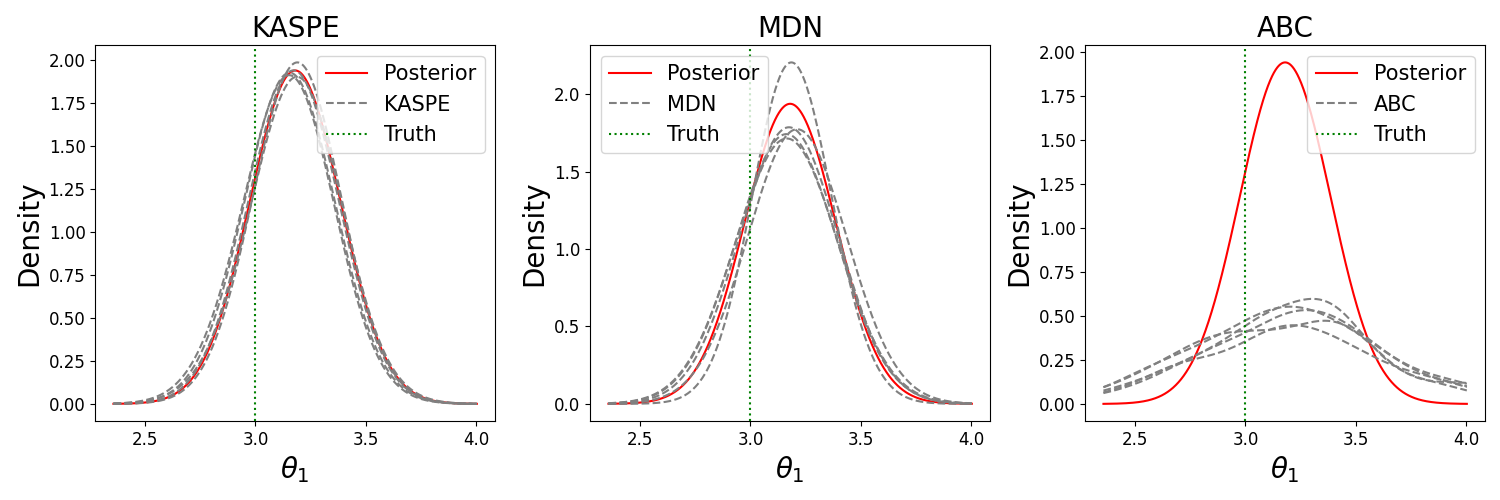}
		}\\
        \vspace{-0.2in}
		\subfigure{%
			
			\includegraphics[width=0.8\textwidth]{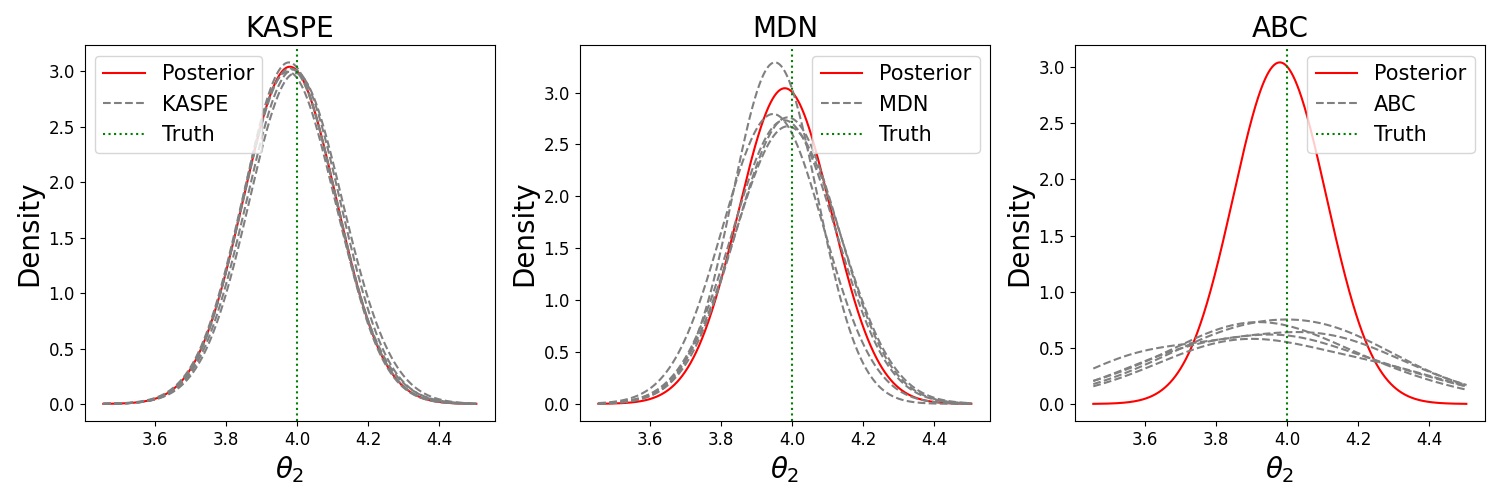}
		}\\
        \vspace{-0.2in}
		\subfigure{%
			
			\includegraphics[width=0.8\textwidth]{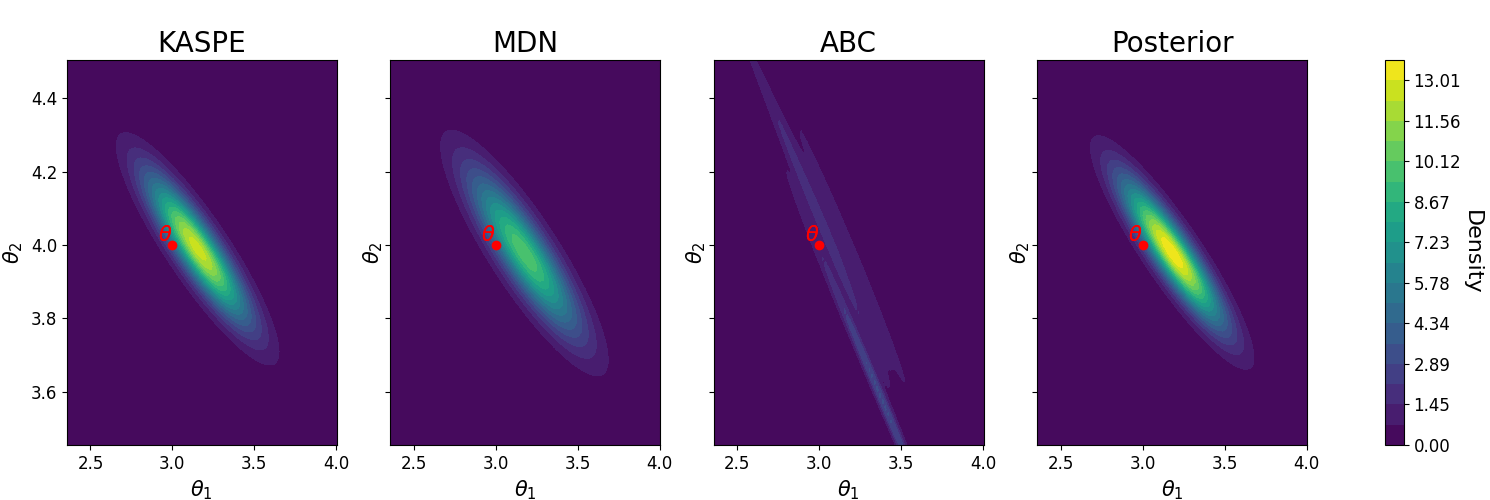}
		}
        \vspace{-0.2in}
	\end{center}
 	\caption{Posterior estimation (marginal densities in the first two rows, joint density in last row) of KASPE, MDN, and ABC (columns), respectively, for the Gaussian mixture model with unknown mean example ($m=100$).}
	\label{Gaussian mixture2}
\end{figure}

\begin{figure}[ht!]
	\begin{center}
		\subfigure{%
			
			\includegraphics[width=0.8\textwidth]{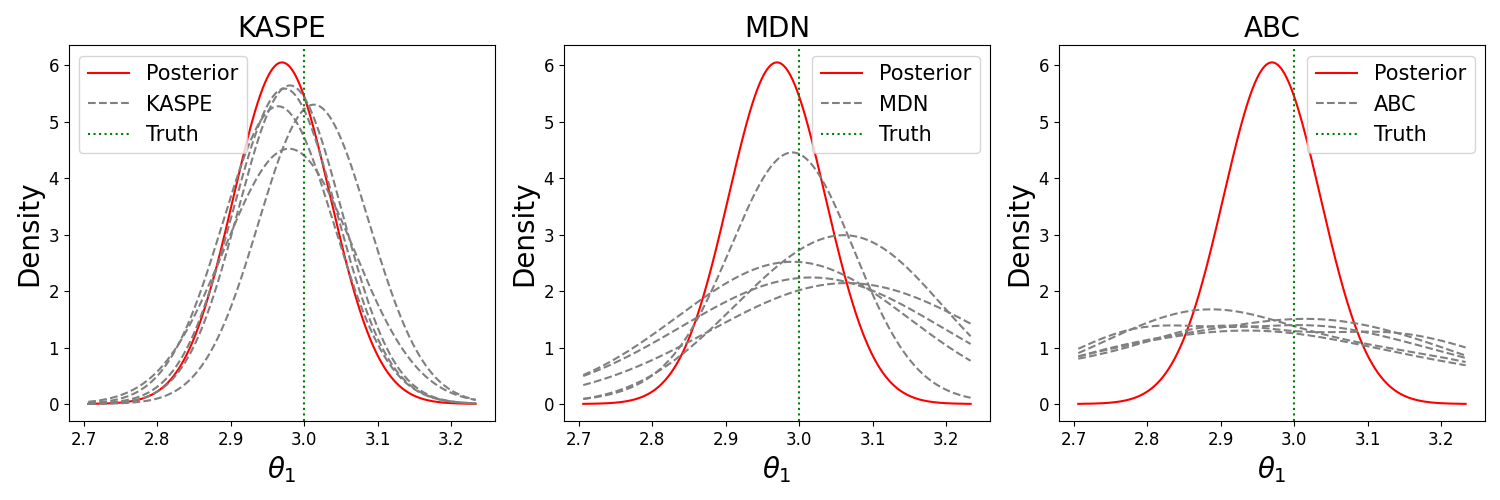}
		}\\
        \vspace{-0.2in}
		\subfigure{%
			
			\includegraphics[width=0.8\textwidth]{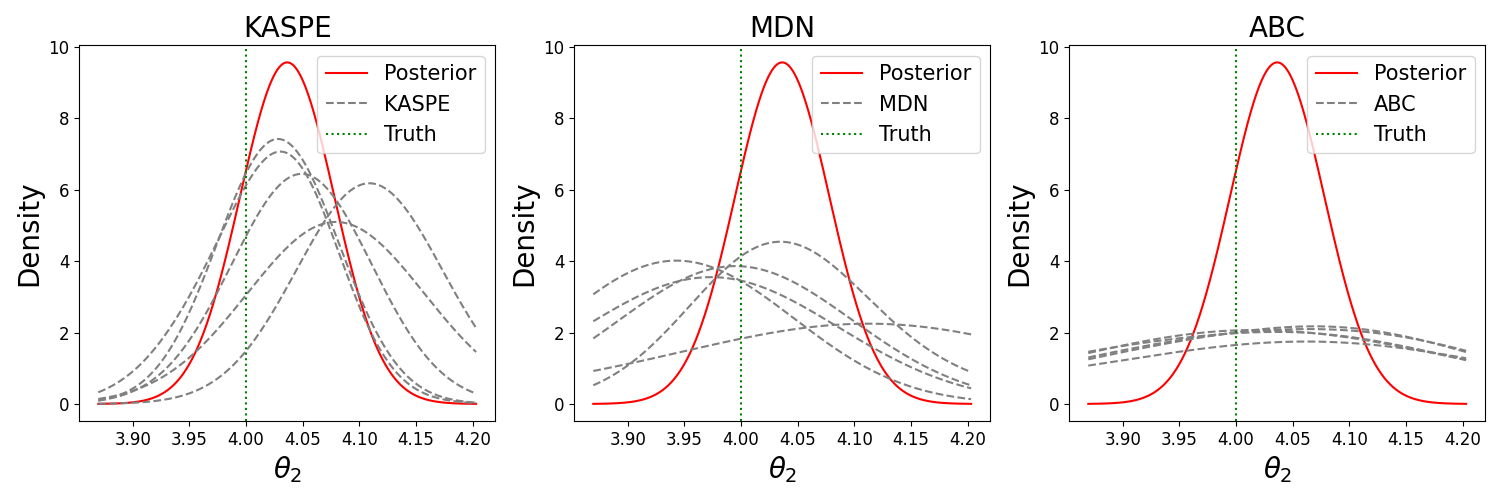}
		}\\
        \vspace{-0.2in}
		\subfigure{%
			
			\includegraphics[width=0.8\textwidth]{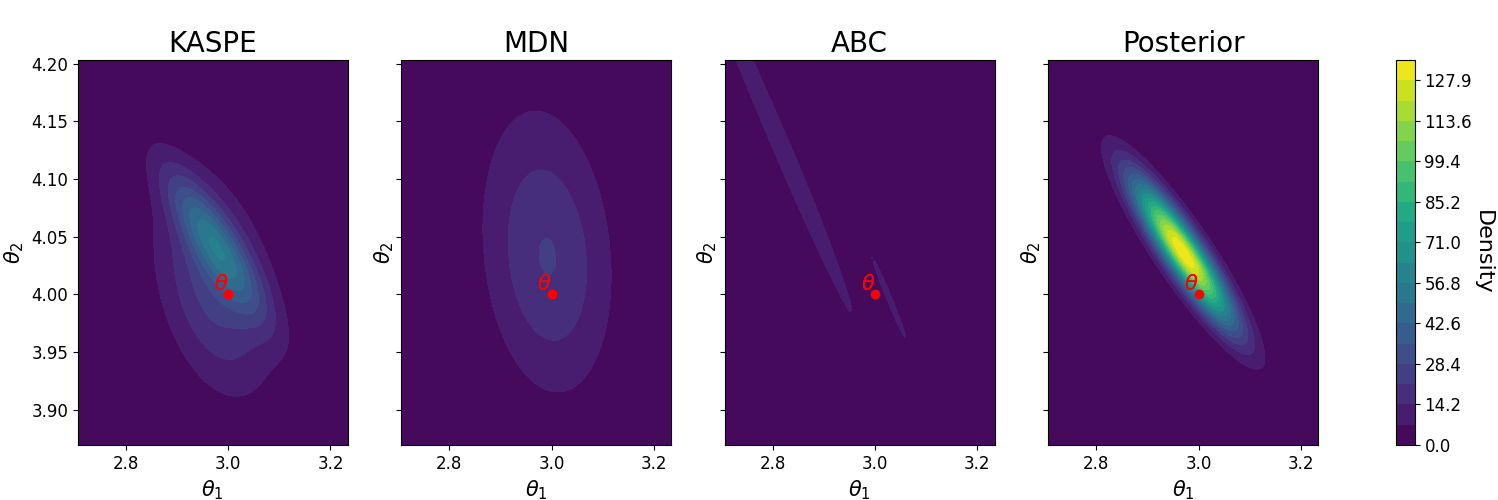}
		}
        \vspace{-0.2in}
	\end{center}
 	\caption{Posterior estimation (marginal densities in the first two rows, joint density in last row) of KASPE, MDN, and ABC (columns), respectively, for the Gaussian mixture model with unknown mean example ($m=1,000$).}
	\label{Gaussian mixture3}
\end{figure}

 \begin{figure}[ht!]
	\begin{center}
	\includegraphics[width=0.9\textwidth]{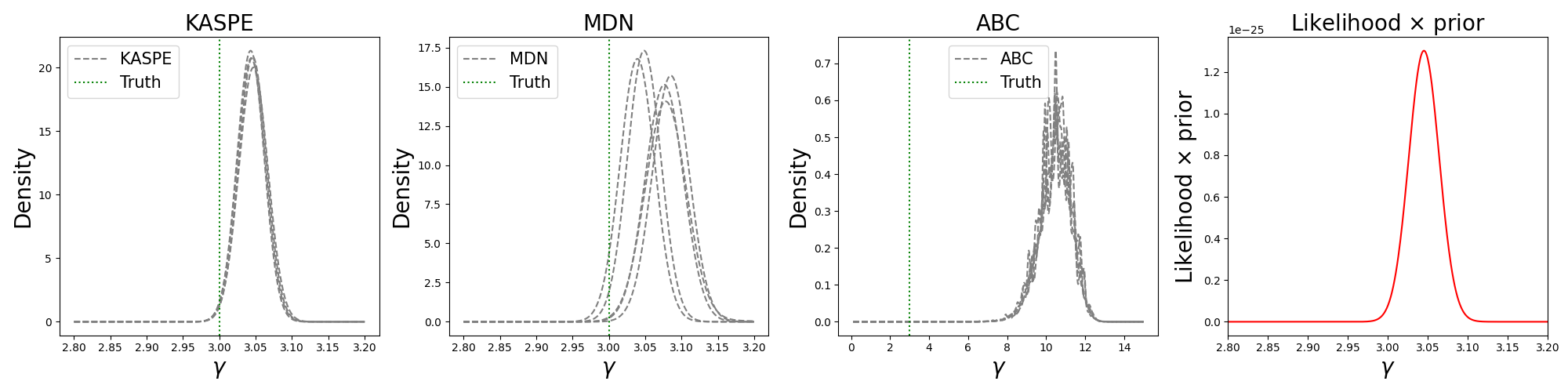}
	\end{center}
    \vspace{-0.2in}
 	\caption{Marginal posterior estimation of KASPE, MDN, and ABC (columns), respectively, for the FN model example ($m=100$).}
	\label{FN2}
\end{figure}

\begin{figure}[ht!]
	\begin{center}
	\includegraphics[width=0.9\textwidth]{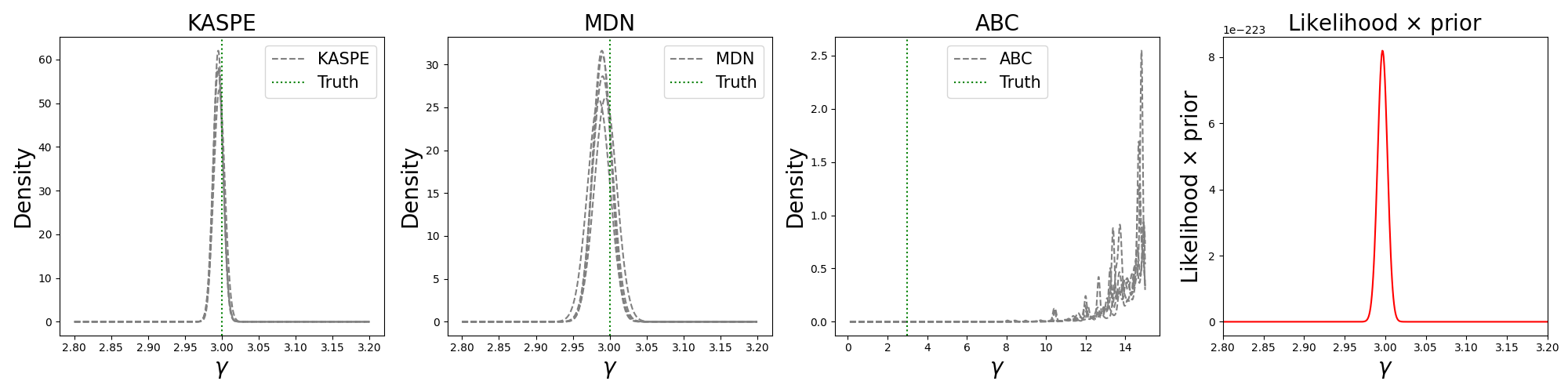}
	\end{center}
    \vspace{-0.2in}
  	\caption{Marginal posterior estimation of KASPE, MDN, and ABC (columns), respectively, for the FN model example ($m=1,000$).}
	\label{FN3}
\end{figure}

\bibliographystyle{agsm}
\bibliography{ref}